\documentclass[fleqn,usenatbib]{mnras}
\usepackage[english]{babel}
\usepackage{graphicx}

\usepackage{amsmath}
\usepackage{amsfonts}
\usepackage{amssymb}
\usepackage{subfig}
\usepackage{comment}
\usepackage{color}
\usepackage{soul}

\usepackage{comment}
\title[Merger events and nuclear activity]{Cosmological simulations
  of black hole growth II: how (in)significant are merger events for
  fuelling nuclear activity?}
\author[Steinborn \& Hirschmann et al.]{Lisa
  K. Steinborn$^{1,2}$\thanks{E-mail: steinborn@usm.lmu.de}, Michaela
  Hirschmann$^{3}$, Klaus Dolag$^{1,4}$, Francesco
  Shankar$^{5}$, \newauthor St\'ephanie Juneau{$^{6,7}$},
   Mirko Krumpe$^{8}$, Rhea-Silvia Remus$^{1}$, Adelheid F. Teklu$^{1,2}$ 
\\
$^{1}$Universit\"ats-Sternwarte M\"unchen, Scheinerstr.1, D-81679 M\"unchen, Germany\\
$^{2}$Excellence Cluster Universe, Boltzmannstra{\ss}e 2, D-85748 Garching, Germany\\
$^{3}$Sorbonnes Universit\'es, UPMC-CNRS, UMR7095, Institut d'Astrophysique de Paris, Boulevard Arago, F-75014 Paris, France\\
$^{4}$Max-Planck-Institut f\"ur Astrophysik, Karl-Schwarzschild
Strasse 1, D-85740 Garching, Germany\\
$^{5}$Department of Physics and Astronomy, University of Southampton, Highfield SO17 1BJ, UK\\
$^{6}$National Optical Astronomy Observatory, 950 N Cherry Ave., Tucson, AZ 85719, USA\\
$^{7}$CEA-Saclay, DSM/IRFU/SAp, F-91191 Gif-sur-Yvette,France\\
$^{8}$Leibniz-Institut f\"ur Astrophysik Potsdam (AIP), An der Sternwarte 16, 14482 Potsdam, Germany}

\begin{document}

\date{Accepted 2018 August 15. Received in original form 2018 May 11}

\pagerange{\pageref{firstpage}--\pageref{lastpage}} \pubyear{2018}

\maketitle

\label{firstpage}

\begin{abstract}
Which mechanism(s) are mainly driving nuclear activity in the centres of galaxies is a major unsettled question. In this study, we investigate the statistical relevance of galaxy mergers for fuelling gas onto the central few kpc of a galaxy, potentially resulting in an active galactic nucleus (AGN). To robustly address that, we employ large-scale cosmological hydrodynamic simulations from the Magneticum Pathfinder set, including models for BH accretion and AGN feedback. Our simulations predict that for luminous AGN ($L_{\rm AGN} > 10^{45} {\rm erg/s}$) at $z = 2$, more than 50 per cent of their host galaxies have experienced a merger in the last 0.5~Gyr. These high merger fractions, however, merely reflect the intrinsically high merger fractions of massive galaxies at $z=2$, in which luminous AGN preferentially occur. Apart from that, our simulations suggest that merger events are not the statistically dominant fuelling mechanism for nuclear activity over a redshift range $z=0-2$: irrespective of AGN luminosity, less than 20 per cent of AGN hosts have on average undergone a recent merger, in agreement with a number of observational studies. The central ISM conditions required for inducing AGN activity can be, but are not necessarily caused by a merger. Despite the statistically minor relevance of mergers, at a given AGN luminosity and stellar mass, the merger fractions of AGN hosts can be by up to three times higher than that of inactive galaxies. Such elevated merger fractions still point towards an intrinsic connection between AGN and mergers, consistent with our traditional expectation.
\end{abstract}


\section{Introduction}
Most, if not all, massive galaxies are nowadays believed to contain a supermassive black hole (BH) in their centers \citep[see e.g.][]{Kormendy, Shankar_2016, Shankar_2017}. During specific, highly variable episodes in the life of a BH, lasting up to $\sim 10^7$~yr, the BH can grow via heavy gas accretion events. Due to resultant gravitational losses, huge amounts of energy can be released, (partly) converted into radiation, causing the BH to shine as an active galactic nucleus (AGN). The required high levels of gas accretion onto a BH demand the supply of gas in the central kiloparsec (kpc) of a BH's host galaxy (fuelling) together with one or more process(es) that make the gas lose its angular momentum, enabling it to move towards the galactic center, i.e. the BH (triggering).   
In general, various processes are believed to be capable of generating the above prerequisites for nuclear activity,
such as: secular evolution bar/disk instabilities (\citealp{Shlosman89});
violently unstable disks (\citealp{Dekel09,  Bournaud11}); gas cooling
from the hot halo (\citealp{Croton06}); galaxy merger events
(\citealp{Silk98, Springel_BHs});  fly-bys
(\citealp{Hopkins_merger}); mass loss from stellar winds
(\citealp{Davies12}); and smooth gas accretion from the halo
(\citealp{King07}) -- in part demonstrated by idealised hydrodynamic simulations of isolated galaxies. However, which of the physical mechanisms are the most efficient and most common drivers for nuclear activity, still remains a heavily debated issue.

Traditionally, merger events have been thought to be the {\it main} process for igniting nuclear activity, simultaneously generating a starburst and forming a stellar bulge in a galaxy. This conventional picture has been historically motivated by the observed relation between the BH and stellar bulge mass (\citealp{Magorrian98, Haering}), by direct observations of
merger signatures in AGN host galaxies \citep[e.g.][]{Sanders}, and
pushed forward by a number of binary merger simulations
\citep[e.g.][]{DiMatteo05, Hopkins_2006, Hopkins_merger}. As a consequence, in many modern semi-analytic galaxy formation models (SAMs), AGN activity is (still) assumed to be mostly driven by major and minor mergers \citep[e.g.][]{DeLucia06, Somerville, Henriques15, Hirschmann16}.

Some theoretical studies, employing either idealised simulations as outlined above, phenomenological models or SAMs, started to challenge this traditional "merger paradigm". Specifically the latter model predictions indicated the necessity to add  other processes as drivers for nuclear activity, in order to reproduce the observed evolution of the AGN luminosity function, in particular the faint end \citep[e.g.][]{Hirschmann_2012, Fanidakis12}. Nevertheless, both refined SAMs as well as phenomenological models point towards an increasing relevance of mergers for driving AGN activity with increasing  luminosity \citep[e.g.,][]{Menci, Hickox_2014, Weigel_2018}.

In addition to theoretical studies, during the last couple of years, an increasingly large amount of observations further severely questioned our
traditional merger paradigm: specifically, \citet{Grogin05,
  Cisternas_2011,   Kocevski_2012,  Villforth_2014,   Rosario_2015,
  Mechtley_2015,   Villforth_2016} find no statistically relevant evidence for an
enhanced fraction of mergers in active galaxies, compared to
a control sample of inactive galaxies  \citep[see,
however,][]{Cotini13, Hong15}.
Even if the majority of modern observational studies agree that for low- and intermediate-luminosity AGN merger events play only a minor role  \citep[see
also][]{DelMoro_2016}, some observations indicate that for luminous
AGN, mergers may still be a statistically relevant driving mechanism, due to measured merger fractions of up to 80~per cent  \citep[][]{Fan_2016,
  Urrutia_2008, Glikman_2015, Treister,   Hopkins_Hernquist_2009}.
In contrast, observations from \citet{Villforth_2016} and \citet{Hewlett_2017} question such a relation:
they find no signs for major mergers being the dominant mechanism for triggering
luminous AGN at $z\sim 0.6$, as their major merger fractions stay
fairly low ($\leq 20\%$); moreover, up to $z=2$, the AGN merger fractions of a given AGN luminosity are only marginally enhanced with respect to those of inactive galaxies.

These rather controversial observational results are likely a consequence of a combination of 
various limitations and complications of AGN surveys: dust obscured AGN/merger signatures; the difficulty in detecting AGN activity delayed relative to the
actual merger event; the visibility of  signatures for (minor) mergers; or other selection effects. As an example, \citet{Kocevski_2015},
\citet{Weston_2016}, \citet{Urrutia_2008}, \citet{Fan_2016}, and \citet{Ricci_2017} find that heavily obscured or reddened  AGN have very high incidences of merger features.
Furthermore, \citet{Juneau_2013} find that galaxies with enhanced specific star formation rates have a higher obscured AGN fraction, which could be linked to an  evolutionary phase in gas-rich mergers.
\citet{Schawinski_2010}, investigating a sample of
early-type galaxies in different evolutionary phases, show that merger
signatures are often hardly visible anymore due to a potentially large
time delay between the merger event and the peak of AGN activity. 
Studies analysing the incidence of nuclear activity with respect to the nearest neighbour separation \citep{Koss10, Ellison_2011, Ellison13, Satyapal_2014} find enhanced fractions of AGN the smaller
the distance to the nearby neighbours (merging galaxy)
and a particularly high AGN fraction in post-mergers,
supporting the time-delay scenario. 
But again, despite this observational evidence that merger events
are {\em principally} capable of driving nuclear activity, most modern studies agree that {\em statistically}, the majority of nuclear activity in AGN populations (dominated by faint and moderately luminous AGN) is likely
 driven by mechanisms other than mergers -- even though many details remain hardly understood.  

To overcome these observational limitations, we can take advantage of
hydrodynamic simulations, which self-consistently capture all
stages of a merger process  and corresponding gas fuelling onto the BH. Up to now, many numerical studies, focusing on AGN driving mechanisms, employed idealised hydrodynamic simulations of isolated galaxies or isolated binary mergers (e.g. \citealp{DiMatteo05, Hopkins_2006,  Hopkins_Quataert_2010,  Capelo}), neglecting any cosmological context and, thus, not following merger fractions and AGN populations over cosmic time. However, recent large-scale {\it cosmological} hydrodynamic simulations (e.g. Magneticum: \citealp{Hirschmann}; EAGLE: \citealp{Schaye}; IllustrisTNG: \citealp{Pillepich18}; MassiveBlack: \citealp{Khandai}; Horizon-AGN: \citealp{Dubois16}; ROMULUS: \citealp{Tremmel_2017}; IllustrisTNG: \citealp{Springel_et_al_2017}), providing statistically relevant and fairly realistic AGN and BH populations \citep[e.g.][]{Hirschmann, Sijacki_2014,  Volonteri_2016, Guevara16, Weinberger_et_al_2017b}, allow us to investigate the {\it statistical  significance} of mergers for  nuclear activity {\it at different cosmic epochs}, 
with respect to other processes, such as smooth gas accretion \citep[][]{Martin_2018}\footnote{Note that the resolution in large-scale cosmological simulations is not high enough to study the role of secular evolution disk instabilities and/or violently unstable disks for nuclear activity or to examine processes driving the gas from the central few kpc to the innermost regions close to the BH.}.
To date, however, a statistical analysis directly linking AGN activity to the merger history and the merger fractions of the host galaxy is still widely lacking.

In this study, we close this gap: we take advantage of the Magneticum
Pathfinder simulation set\footnote{www.magneticum.org} (Dolag et
al. in prep., \citealt{Hirschmann}) to statistically investigate the role of
merger events for driving nuclear activity in a galaxy. Due to limited resolution in a cosmological set-up, our analysis is restricted to explore {\it the impact of mergers on fuelling gas onto the central few kpc} of a galaxy. 
Specifically, our analysis evolves around two related questions: 
\begin{itemize}\vspace{-0.2cm}
\item To what extent does the merger history affect the
  incidence for nuclear activity in galaxies, as well as the ISM properties in the 
  central few kpc of a galaxy, which are controlling the accretion luminosities? 
  \item What is the probability that an AGN host galaxy of a given luminosity has
  experienced a recent merger event, and to what extent do these
  merger fractions reflect an intrinsic AGN-merger connection?
  \end{itemize}\vspace{-0.2cm}
  This paper is the second in a series focusing on BH growth and
 AGN populations using the Magneticum set. The simulations are
 described  in detail in Paper I \citep{Hirschmann}, which
 demonstrated that AGN luminosities together with their
 anti-hierarchical trend are consistent with observations over cosmic
 time. In addition, our simulations can successfully reproduce various
 other observed galaxy and BH properties
 \citep[e.g.,][]{Teklu_2015,
   Steinborn_2015, Remus17, Teklu_2017, Remus_2017b, Teklu_2018, Schulze_2018} providing an ideal testbed for our study. We emphasize that thanks to the uniquely large simulated volume of (500 Mpc)$^3$, we are able to study  the AGN-merger connection for the
 rarest very luminous quasars, which are not accessible in most other
 state-of-art simulations (like EAGLE, Illustris). 
   
This study is structured as follows. We briefly describe the
simulation details as well as the algorithm for identifying merger
events in Section \ref{simulations}. In Section \ref{results}, we
first analyse AGN light curves for five test cases and connect them to
the recent merger history (Section \ref{testcases}); then we  turn
to the full AGN population and explore the statistical role of mergers
for fuelling nuclear activity (Section \ref{populationstudy}). To
understand the physical origin of our results, in Section
\ref{mergerrates_host}, we focus on the impact of galaxy properties on the BH accretion in our
simulations. We discuss our findings in the context of previous
theoretical and observational studies and address possible caveats of
our method in Section \ref{discussion}. Finally, Section
\ref{conclusion} summarizes our results.

\section{The Magneticum Pathfinder Simulations}
\label{simulations}

\subsection{The theoretical set-up}
For this analysis, we employ two cosmological, hydrodynamic
simulations taken from the set of the Magneticum Pathfinder
Simulations performed with the TreePM-SPH code \textsc{p-gadget3} (a follow-up version of \textsc{gadget2}, \citealt{Springel}), including isotropic thermal conduction
\citep{Dolag04} with an efficiency of $\kappa=1/20$ of the classical
{\it Spitzer} value \citep{Arth}.  
These simulations assume the currently favoured standard $\Lambda$CDM
cosmology, where the Hubble parameter is $h=0.704$ and the density 
parameters for matter, dark energy and baryons are
$\Omega_\mathrm{m}=0.272$, $\Omega_\mathrm{\Lambda}=0.728$ and
$\Omega_\mathrm{b}=0.0451$, respectively (WMAP7,
\citealt{Komatsu}\footnote{Note that changing to the more recent
  Planck cosmological parameters is not expected to significantly
  affect our results.}). 
The simulation code includes effective models for different baryonic
processes such as star formation \citep{Springel_Hernquist}, stellar
evolution, metal enrichment, and supernova feedback
(\citealt{Tornatore03}, \citealt{Tornatore}) as well as a radiative
cooling and photo-ionization heating due to a constant UV
background. The net cooling function depends on the individual metal
species following \citet{Wiersma}. 
Most importantly, our code accounts for the growth of  BHs and their associated AGN feedback. The BH accretion rate is
computed based on the Bondi model \citep{Hoyle,   Bondi, Bondi_Hoyle}
as presented in \citet{Springel_BHs}: 
\begin{equation}
\dot{M}_\mathrm{B} = \frac{4 \pi \alpha G^2 M_\bullet^2
  \rho_{\mathrm{gas}}}{(v_{\mathrm{rel}}^2 + c_{\mathrm{s}}^2)^{3/2}}, 
\label{accretion_rate}
\end{equation}
where $\alpha=100$ is an artificial boost factor \citep{Springel_BHs},
$<\rho_{\mathrm{gas}}>$ is the mean density, $<c_s>$ the mean sound
speed, and $<v_{\mathrm{rel}}>$ the mean velocity of the gas in the
resolved accretion region relative to that of the BH.
Note that sub-kpc accretion flows onto the BHs as well as the Bondi accretion radius
are unresolved in large-scale cosmological simulations.
Therefore, we can only capture BH growth due to the larger scale
gas distribution within the ``numerically'' resolved accretion region
$r_\mathrm{acc}$, which is defined by a specific number of neighbouring particles (distance to the 295th neighbour).

To model feedback from the accretion onto the BH, we assume an isotropic
thermal energy release into the ambient gaseous medium following
\citet{Springel_BHs} with the modifications from \citet{Fabjan}, where
the AGN feedback efficiency for radiatively inefficient AGN is
increased in order to mimic observed inflated hot bubbles in radio
galaxies. Note that the change of BH accretion rates with resolution (because of different central gas properties and accretion radii) is compensated by adjusting the feedback efficiency such that BH masses are always consistent with the observed BH scaling relations. To what extent our analysis depends on the specific BH accretion and AGN feedback schemes adopted in our simulation, will be discussed in Section \ref{discussion}.  For further simulation and model details, we refer the
reader to Paper I.  

In the course of this paper, we analyse the following two cosmological simulations from the Magneticum simulation set:
\begin{itemize}\vspace{-0.2cm}
\item{{\bf 68Mpc/uhr:}
        This simulation has a volume of $(68
        \mathrm{Mpc})^3$ combined with a comparably high resolution, with dark
        matter and gas particles masses of $M_{\mathrm{dm}}=3.7
        \cdot 10^7 M_{\odot}/h$ and $M_{\mathrm{gas}}=7.3 \cdot 10^6
        M_{\odot}/h$, respectively. This resolution is high enough to
        largely capture the internal structure and morphology of
        galaxies \citep{Teklu_2015, Teklu_2017}.
        Note that BH accretion rates in the intrinsic, code-based time resolution are stored only at $z \geq 1.5$, allowing us to study detailed BH light curves down to that redshift (Fig. \ref{lightcurves}).
        }
\item{{\bf 500Mpc/hr:}
       The second simulation considered in this work comprises a large volume of $(500 
        \mathrm{Mpc})^3$ with a resolution of $M_{\mathrm{dm}}=6.9
        \cdot 10^8 M_{\odot}/h$ and $M_{\mathrm{gas}}=1.4 \cdot 10^8
        M_{\odot}/h$, enabling us to study the evolution of a large AGN population, including} very massive and very luminous AGN \citep[][]{Hirschmann}. 
        This simulation run is publicly available via our web
        interface \citep[see][]{Ragagnin_2017}.
\end{itemize}\vspace{-0.2cm}

The two simulations are performed with the same settings in terms of
physical processes and cosmology, but cover different mass ranges due
to different box sizes and resolutions.
The 68Mpc/uhr simulation is solely used to study individual AGN light curves of five test  cases (Section \ref{testcases}). In the remainder of the paper, we show the results for the 500Mpc/hr simulation due to its better statistics.
For this simulation run we consider only galaxies
above a certain resolution threshold of
$M_*>10^{11}M_{\odot}$
(corresponding to a particle number of roughly 2800 particles)\footnote{Since we trace the galaxies back in time, the
  progenitor galaxies can have much smaller masses, especially the
  less massive galaxies in minor mergers. Therefore, the resolution
  limit is chosen to be fairly conservative.}.
We have
explicitly verified that the results  qualitatively\footnote{Note that a direct comparison between the two simulations is not possible due to the different mass ranges.} converge towards higher
resolution.

\subsection{Halo identification and merger tree construction}
The simulation predictions are output in 145 snapshots with equal 
time intervals between the snapshots\footnote{ For redshifts $z>1$
  the simulation output has larger time intervals than for $z<1$.}. 
For each snapshot, haloes and subhaloes are identified using the
friends-of-friends algorithm \citep[FOF,][]{Davies_1985} assuming a
linking length of 0.16 in combination with \textsc{subfind} (\citealt{Dolag09}, 
\citealt{Springel_subfind}).  

We continue to connect haloes and subhaloes over time, i.e. we
construct merger trees using the \textsc{L-HaloTree} algorithm, which is
described in the supplementary information of \citet{Springel_Millenium}.
In short, to determine the appropriate descendant, the unique IDs that
label each particle are tracked between outputs. For a given halo, the
algorithm finds all haloes in the subsequent output that contain some
of its particles. These are then counted in a weighted fashion, giving
higher weight to particles that are more tightly bound in the halo
under consideration. The weight of each particle is given by
$(1+j)^{-\alpha}$, where $j$ is the rank, based on its binding energy,
as returned by \textsc{subfind}, and $\alpha$ is typically set to $2/3$. This
way, preference is given to tracking the fate of the inner parts of a
structure, which may survive for a long time upon in-fall into a
bigger halo, even though much of the mass in the outer parts can be
quickly stripped. Once these weighted counts are determined for each
potential descendant, the one with the highest count is selected as
the descendant. Additionally, the number of progenitors is calculated
for each possible descendant. \textsc{L-HaloTree} is constructing descendants
(and its associated progenitors) for $A\rightarrow B$ as well as
$A\rightarrow C$. Therefore, as an additional refinement, some haloes
are allowed to skip one snapshot $B$ in finding a descendant, if
either there is a descendant found in $C$ but none found in $B$, or,
if the descendant in $B$ has several progenitors and the descendant in
$C$ has only one. This deals with cases where the algorithm would
otherwise lose track of a structure that temporarily fluctuates below
the detection threshold. 

In this approach, two galaxies are defined to have merged, as soon as 
they are identified as only one galaxy by \textsc{subfind}, i.e. as soon as
they are gravitationally bound to each other. For the following 
analysis, we connect an AGN with a merger signature of its host galaxy, 
if a merger happened up to 0.5~Gyr before the time step the AGN luminosity is computed.
The time interval of maximum 0.5~Gyr is motivated by our case studies in Section \ref{testcases} showing that mergers have hardly any effect onto the AGN activity after 0.5~Gyr. This is supported by previous simulations of isolated galaxy mergers \citep[e.g.][]{Hopkins_merger, Johansson_2009}, also finding no significant effect on the AGN activity more than 0.5~Gyr after the merger event. It is also unlikely that merger signatures would be visible/detectable in observations after such a time period. But note that we explicitly tested larger time intervals up to 1.5~Gyr, without finding any qualitative difference in our results.

Throughout this analysis, once merger events have been identified, we divide our galaxies and AGN hosts into 
three different ``merger'' classes, depending on the stellar merger
mass-ratio $M_{*2}/M_{*1}$ ($M_{*1}$ and $M_{*2}$ are the stellar
masses of the more and less massive progenitor galaxy, respectively): 
\begin{itemize}
    \item{{\bf no mergers}, including very minor mergers with
        $M_{*2}/M_{*1}<1:10$,} 
    \item{{\bf at least one minor merger:} $1:10<M_{*2}/M_{*1}<1:4$ (but no major mergers)}
    \item{{\bf at least one major merger:} $M_{*2}/M_{*1}>1:4$ (eventually additional minor mergers).}
\end{itemize}
Note that, if a galaxy/AGN host has experienced both major and minor mergers during the last 0.5~Gyr, it is added to the major merger class (due to the common believe that major mergers are more significant for nuclear activity than minor mergers).

Such a division into different merger classes is further complicated by defining/deriving mass-ratios for  merger events, potentially affected by artificial false estimations of the stellar merger mass-ratios, as a consequence of the \textsc{subfind} algorithm. In fact, the physically most meaningful estimation of the merger mass-ratio is not necessarily made at the time when the merger is identified by \textsc{subfind}, since at that time in-falling galaxies can have already suffered from tidal stripping and other environmental effects (distracting the intrinsic mass-ratio). In order to circumvent such problems, we consider the {\it maximum} stellar mass-ratio between two merging galaxies during the past 1.5~Gyr. In the appendix, we describe our merger identification algorithm and the estimation of the stellar merger mass-ratio in more detail.

%%%%%%%%%%%%%%%%%%%%%%%%%%%%%%%%%%%%%%%%%
\section{Relation between merger events and nuclear activity}
\label{results}
%%%%%%%%%%%%%%%%%%%%%%%%%%%%%%%%%%%%%%%%%

In this section, we investigate  to what degree nuclear activity of a galaxy is related to its recent merger history. We remind
the reader that due to limited resolution in state-of-the-art large-scale cosmological simulations (including the Magneticum simulations considered here), BH accretion is governed by ISM properties (density, temperature and 
relative velocity) in the central few kiloparsec of a galaxy, following the Bondi-Hoyle approach (equation \ref{accretion_rate}). Thus, by construction, we are only able to
examine {\em the impact of merger events on fuelling the gas onto the central few kpc, and on the correspondingly estimated nuclear activity}. 

We first consider five representative test cases of AGN galaxies above $z=2$ from the 68Mpc/uhr simulation, individually discussing their AGN light curves with respect to the underlying merger history (subsection \ref{testcases}). Turning to the full AGN population, as predicted by the 500Mpc/hr simulation run (subsection \ref{populationstudy}), we analyse the statistical incidence for nuclear activity in galaxies as a function of their merger history and the AGN
luminosity. We further quantify the maximum probability for AGN to be potentially fuelled by mergers by computing the merger fractions of AGN host galaxies, confronting them with observational estimates.
Note that throughout this study, bolometric AGN luminosities are
calculated from the BH accretion rates following \citet{Hirschmann}.

%%%%%%%%%%%%%%%%%%%%%%%%%%%%%%%%%%%%%%%%%
\subsection{Five case studies}\label{testcases}
%%%%%%%%%%%%%%%%%%%%%%%%%%%%%%%%%%%%%%%%%

\begin{figure*}
  \includegraphics[trim = 1mm 1mm 1mm 14mm, clip, width=0.9\textwidth]{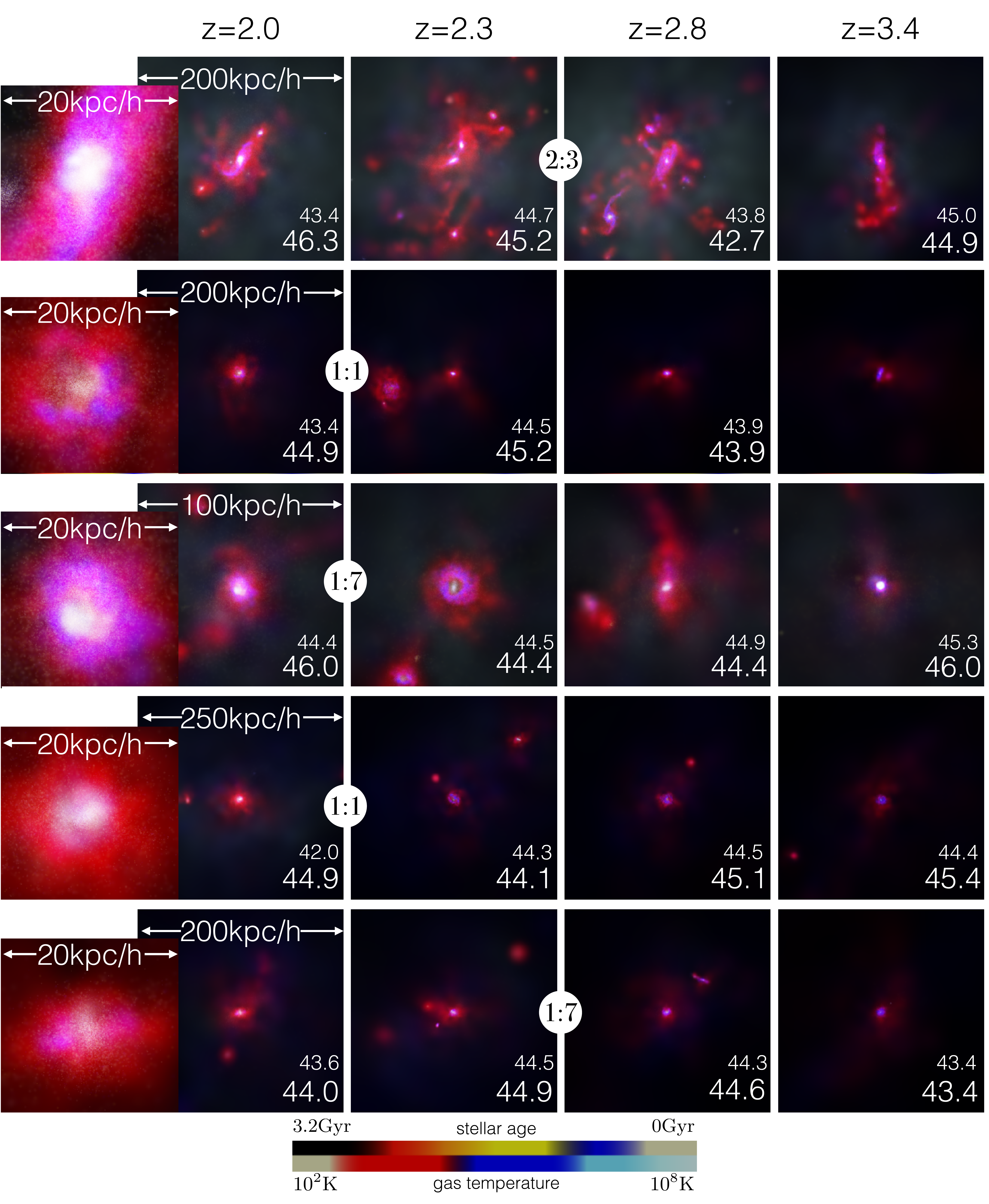}
  \caption{The different columns visualize the gaseous and stellar
    component (colour-coded by the gas temperature and the stellar
    age, respectively, as indicated by the colour-bar) of five
    different merging galaxies (different rows) in the 68Mpc/uhr
    simulation at $z=2.0$ (left-hand panels), and their progenitors
    0.5~Gyr, 1.0~Gyr, and 1.5~Gyr before $z=2$ (columns towards the
    right-hand side). When the galaxies host a SMBH, its instantaneous luminosity
    (log($L$) in [erg/s]) value at the given snapshot is specified by the large numbers in the bottom right of each
    panel.
    The small numbers above the large ones are the average AGN luminosity values within a timeintervall $z \pm 0.01$ around the given snapshots.
    The white circles and their numbers
    indicate the stellar merger mass ratio and their positions
    correspond to the time at which the merger mass ratio
    has been determined. 
	  }
\label{collage}
\end{figure*}
\begin{figure*}
  \includegraphics[trim = 10mm 5mm 5mm 10mm, clip, width=0.9\textwidth]{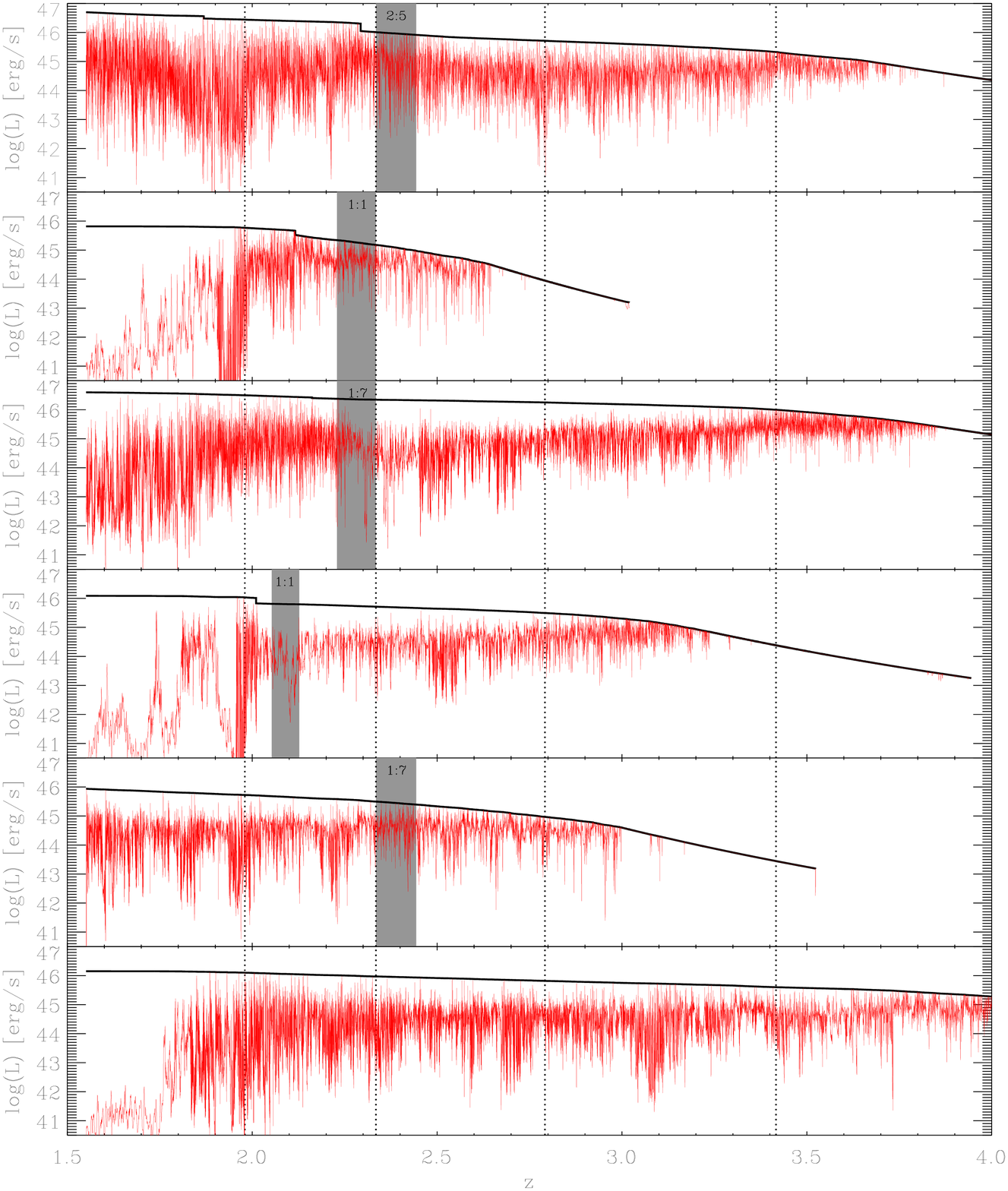}
  \caption{Red lines show the light curves for the examples shown in
    Fig. \ref{collage} (row 1-5, same order) as well as for one additional example without a recent merger (bottom row). The black solid lines show the
    Eddington luminosity, i.e. the maximum luminosity allowed in the
    simulation. 
  Black dotted lines indicate the snapshots taken from the
  simulation. 
  The grey shaded areas show the redshift range within which the
  merger has been identified. The corresponding merger mass ratio is
  given in the top of these areas. 
	  }
\label{lightcurves}
\end{figure*}

%%%%%%%%%%%%%%%%%%%%%%%%%%%%%%%%%%%%%%%%%
\subsubsection{The evolutionary sequence of AGN  hosts at $z=2$}
%%%%%%%%%%%%%%%%%%%%%%%%%%%%%%%%%%%%%%%%%

We start with investigating the AGN-merger connection by
selecting five different example AGN hosts at $z=2$, having experienced a major
or minor merger event in the past 1~Gyr, i.e. between $z=2.0$ and $z=2.8$. Fig. \ref{collage} visualises the  gaseous and stellar distributions (colour-coded as
indicated by the colour bar)\footnote{performed with the free software
  Splotch, http://www.mpa-garching.mpg.de/$\sim$kdolag/Splotch from   
  \citet{Dolag_splotch}} of the five example galaxies at $z=2$ (first and second columns) and that of their progenitors at $z=2.3$, $z=2.8$, and $z=3.4$ (third, fourth and fifth column, respectively). The stellar merger mass-ratios ($M_{*2}/M_{*1}$) are shown by the
white circles, whose positions indicate at which time the merger mass-ratio has been computed. In all cases, merger signatures such as
tidal tails are still visible at $z=2$. The instantaneous AGN luminosities (log($L_{\mathrm{bol}}$) in [erg/s]), affected by AGN variability,
 are specified in the bottom right of each panel\footnote{In the right panel in the second row, no BH luminosity has been specified, since the BH has not yet been seeded.},
while the small numbers above indicate the logarithmic average AGN luminosity in a time interval of $z \pm 0.01$ before and after the time of the snapshot. In four out of five examples the instantaneous luminosity is larger in the snapshot after the merger than in the one before the merger, for some AGN, however, only slightly. The average AGN luminosity, instead, increases only in two out of the five test cases. But even if the average values do represent the global trends more accurately, the instantaneous luminosities better reflect observations.  Indeed, in our simulations most AGN are only rather short outbursts due to the large AGN variability, as we shall see in the next section. 

The first row in Fig. \ref{collage} shows two gas-rich spiral-like galaxies, which merge between $z=2.8$ and $z=2.3$. Between these redshifts, the instantaneous and average AGN luminosity increases by 2.5~dex and 0.9~dex, respectively, indicating that the merger boosts the accretion onto the central BH.
In the second row, a 1:1 merger is identified between $z=2.3$ and $z=2.0$, but the AGN luminosity slightly decreases.
The third row illustrates a minor merger of two gas-rich galaxies. Although the merger mass-ratio is much smaller than in the first example, the AGN luminosity
increases significantly, from $\log(L_{\mathrm bol}) = 44$~erg/s to
$46$~erg/s.
This is, however, not reflected in the average luminosity values.
The last two examples show the evolutionary sequence of
two moderately luminous AGN whose host galaxies have experienced a major 
(fourth row) and a minor (fifth row) merger.
During the merger shown in the fourth row, the average AGN luminosity strongly decreases by more than 2~dex, while the instantanoues value is larger after the merger than before the merger.
In the example in the fifth row, the luminosity is hardly changing during the merger.
Thus, the five examples shown in Fig. \ref{collage} suggest that merger events {\it may}, but do not {\it necessarily} boost the accretion onto BHs.

%%%%%%%%%%%%%%%%%%%%%%%%%%%%%%%%%%%%%%%%%
\subsubsection{AGN light curves}
%%%%%%%%%%%%%%%%%%%%%%%%%%%%%%%%%%%%%%%%%

For a deeper understanding of the inconclusive AGN-merger connection seen so far, Fig. \ref{lightcurves} explicitly illustrates the AGN light curves of our five example galaxies from $z=4$ down to $z=1.5$ as well as, for reference, of one example AGN without a recent merger (bottom row). Note that, while the simulation code stores BH accretion rates also between two snapshots, i.e. for very  small time steps of $\sim 0.1$Myr, the host galaxy properties are only  accessible at the time of the snapshots (i.e. with larger time steps). The simulation snapshots (as depicted in Fig. \ref{collage}) are indicated by the black dotted lines in Fig. \ref{lightcurves}. The times during which the mergers have been identified in the simulation are marked as grey shaded areas, with the merger mass-ratio indicated in the top of these areas.   

The first five light curves in Fig. \ref{lightcurves} illustrate that right after the seeding of the BH in a galaxy, the BH accretes gas at rates close to or at the Eddington limit, which are by default the maximum luminosity allowed in the simulation (black  solid line in  Fig. \ref{lightcurves}). During that phase, the accretion rates are likely artificially high due to our low BH seeding mass relative to the galaxy stellar  mass\footnote{See Fig. 4 and the corresponding discussion in \citet{Steinborn_2015} for further details.}. After this first accretion phase at or close to the Eddington-limit, AGN luminosities become highly variable over smallest time steps of $\sim 0.1$Myr. 

In the top, second and fifth panel of Fig. \ref{lightcurves}, both minor and major mergers increase or decrease the AGN luminosity {\it only marginally}. In these cases, already before the merger, the BH can accrete at/close to the Eddington limit, due to large amounts of gas available at these early times, so that a merger does not have any significant, additional effect.
As the amount of gas in galaxies varies with redshift, this may imply that the relevance of mergers for nuclear activity is also dependent on redshift.
Despite the higher BH mass after the BH merger, resulting in a higher Eddington limit, and thus, higher maximum AGN luminosity (solid black line), the high AGN variability leads to an AGN luminosity at $z=2$ not being necessarily larger at the time of "observation" (i.e. when the snapshot is written) than before the merger and can, in fact, also be lower (see, e.g., the second panel in Fig. \ref{lightcurves}).

The light curve for a merger-free AGN in the bottom panel of Fig. \ref{lightcurves} additionally shows that similarly high AGN luminosities as seen in the top, second and fifth panel can also be induced by processes other than mergers. Interestingly, similar to  our two test cases of 1:1 mergers (second and fourth panel), the AGN activity declines rapidly from $z=2$ to $z=1.5$, possibly either due to starvation or due to disturbances of the morphology and/or the dynamics of the gas within the central kpc.\footnote{We verified that there is also no merger between $z=1.5$ and $z=2$.}

In the light curves shown in the third and the fourth panels of Fig. \ref{lightcurves}, the average AGN luminosity significantly rises during and right after the merger event. In both cases, the BH accretes at fairly low Eddington-ratios before the merger, while after the merger, the BH accretion can reach the Eddington limit. This seems to suggest that a merger is more likely to boost AGN luminosity, if the BH was rather inactive before the merger (due to low amount of gas, missing gas inflows etc.).

To summarize, our five case studies demonstrate that analysing the effect of merger events on nuclear activity is significantly complicated by strong variations in  the evolution of BH accretion rates. The net increase or decrease in AGN luminosity between the times of two snapshots, (see Fig. \ref{collage}), is distorted by the significant flickering in AGN luminosity:  considering the AGN luminosity only at a specific time of one of our snapshots (dashed black lines) does not necessarily reflect the {\em average} AGN luminosity in a representative way (but note, this is the same for observations). 

To still find a meaningful connection between the nuclear activity and the merger history of the host in our simulations, we can either average over the BH accretion rates of a galaxy within a given time interval (centred at the time of the snapshot), or we can investigate the AGN luminosities of a {\it statistically large sample} at a given time-step. In this study, we follow the latter approach. Nevertheless, we verified that an additional averaging over the AGN luminosities within a given time interval does not affect our results, even when restricting to the most luminous AGN.

%%%%%%%%%%%%%%%%%%%%%%%%%%%%%%%%%%%%%%%
\subsection{AGN population study}\label{populationstudy}
%%%%%%%%%%%%%%%%%%%%%%%%%%%%%%%%%%%%%%%

The results presented for the five AGN test cases raise the
questions, (i) how frequently mergers increase AGN activity on
a statistical basis and (ii) to what extent such a boost is dependent on AGN
luminosity or the merger mass-ratio. To ensure sufficiently high statistics, in this section we consider large populations of AGN and their host galaxies in the 500Mpc/hr run of the Magneticum set. First, we examine the statistical incidence
of nuclear activity in galaxies as a function of their recent merger
history, giving us the maximum probability that a merger event can
fuel nuclear activity in galaxy populations (subsection \ref{incidence}). Then, we quantify the maximum likelihood that nuclear activity in AGN populations can be merger-induced (subsection \ref{mergerrates}) and their dependence on AGN luminosity, also compared to observations (subsection \ref{mergerrates_lum}).  Further comparing  merger fractions of AGN hosts to that of inactive galaxies allows us to assess to what extent merger events are  actual drivers for nuclear activity.

%%%%%%%%%%%%%%%%%%%%%%%%%%%%%%%%%%%%%%%
\subsubsection{Incidence for nuclear activity in galaxies as a function of their merger history}\label{incidence}  
%%%%%%%%%%%%%%%%%%%%%%%%%%%%%%%%%%%%%%%

\begin{figure}
  \includegraphics[trim = 0mm 10mm 0mm 17mm, clip,
  width=0.5\textwidth]{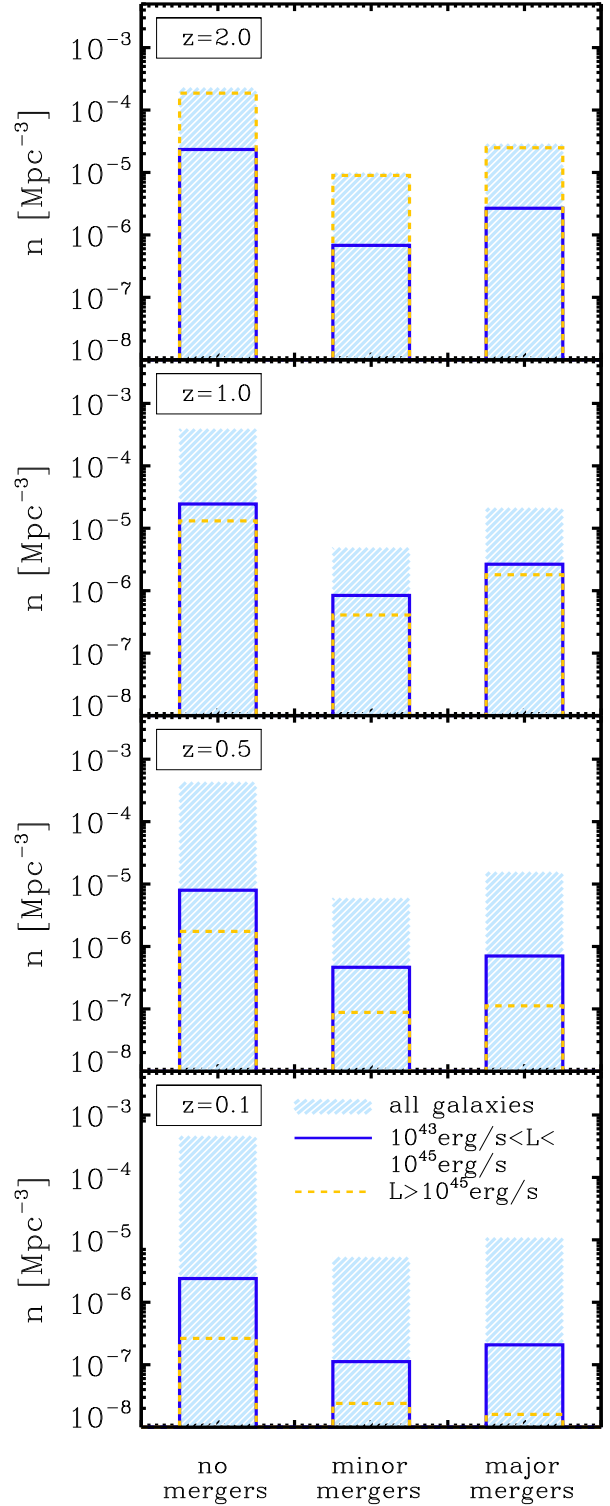} 
  \caption{Number density of all galaxies (hatched areas), moderately luminous AGN ($10^{43}$erg/s$<L<10^{45}$erg/s; solid blue lines), and luminous AGN ($L_{\mathrm{bol}}10^{45}$~erg/s; dashed yellow lines). We include only galaxies with stellar masses above our resolution limit ($M_*>10^{11} M_{\odot}$) and distinguish between galaxies which have experienced no mergers (including very minor mergers with $M_{*2}/M_{*1}<1:10$), minor mergers ($1:10<M_{*2}/M_{*1}<1:4$), and major mergers ($M_{*2}/M_{*1}>1:4$) in the past 0.5~Gyr at $z=2.0,1.0,0.5,0.1$ (panels from top to bottom).  
	  }
\label{merger_mass_ratio}
\end{figure}

Fig. \ref{merger_mass_ratio}  shows the number density of all galaxies
(light blue hatched area), of moderately luminous and luminous active galaxies with $10^{43}$erg/s$<L_{\rm bol}<10^{45}$erg/s and $L_{\rm bol}>10^{45}$erg/s (blue solid and yellow dashed lines), respectively, having
experienced either no mergers (left bar), minor (middle bar) or major
mergers (right bar) in the last 0.5~Gyr at $z=2, 1, 0.5, 0$ (panels
from top to bottom).
As expected from a hierarchical structure formation scenario, the number density of all galaxies with major or minor mergers is decreasing from z=2 to z=0.
Over the same redshift range, the number density of all galaxies {\it without} recent mergers is marginally increasing.

Instead, the number density of AGN {\em always} decreases from $z=2$ to $z=0$, also for host galaxies {\em without} a recent merger. The more luminous AGN are, the stronger AGN number densities are declining towards lower redshift. While at $z=2$ nearly all galaxies with a recent merger event host a luminous AGN, at $z=0.1$, it is only a small fraction of less than 10 per cent for moderately luminous and of less than 1 per cent for luminous AGN.  

\begin{figure}
  \includegraphics[trim = 0mm 7mm 3mm 11mm, clip, width=0.5\textwidth]{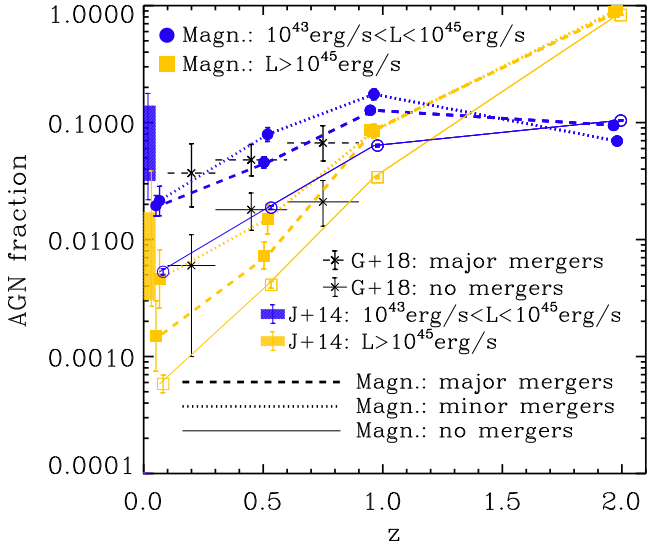}
  \caption{Redshift evolution of  fractions of AGN host galaxies
    with  $10^{43}$erg/s$<L<10^{45}$erg/s (blue circles) and
    $L_{\mathrm{bol}}>10^{45}$erg/s (yellow squares) having
    experienced a major (dashed lines), minor (dotted lines) or no
    merger (solid lines) in the past 0.5~Gyr assuming a stellar mass
    cut of $M_*>10^{11} M_{\odot}$. Error bars in the AGN fractions
    indicate the binomial confidence intervals.
    The blue and yellow bars show observed fractions of local AGN with $10^{43}$erg/s$<L<10^{45}$erg/s and $L>10^{45}$erg/s, respectively \citep[][]{Juneau_2014}.
    The black crosses with the error bars show the observed datapoints from \citet[][dashed and solid bars indicate the redshift range of galaxies with major mergers and galaxies without any merger signatures, respectively]{Goulding_2018}.
    }
\label{AGN_fraction_z}
\end{figure}

Fig. \ref{AGN_fraction_z} further quantifies such AGN fractions: shown is the redshift evolution of the {\it ratio}  of the number (density) of moderately luminous and luminous AGN hosts (blue circles/lines and yellow squares/lines, respectively) to that of all
galaxies (i.e., the AGN duty cycle), having experienced in the past 0.5~Gyr either no mergers (solid lines),
i.e. $N_{\rm AGN,\ no\ merger}/N_{\rm all,\ no\ merger}$, or minor/major mergers (dotted/dashed lines), i.e. $N_{\rm AGN,\ minor/major}/N_{\rm all,\ minor/major}$.
The error-bars indicate the binomial confidence intervals. 

At $z=2$ almost 100~per cent of all galaxies host an AGN, and more than 90~per cent even a luminous AGN  ($>10^{45}$~erg/s), irrespective of the recent merger history. This result implies that mergers do not necessarily play any role for nuclear activity at these early times: large amounts of turbulent gas available in and around such young galaxies can lead to radial gas inflows onto the central few kpc, and thus, to high accretion rates onto BHs, also without any recent merger event.   

Towards lower redshifts ($z<2$), the situation changes: independent of the
recent merger history, the fractions of luminous AGN are strongly declining to less than 1 per cent at z=0.1, as a consequence of the generally reduced gas content and density in the inner region of a galaxy (see Section \ref{gas_properties} for further discussion). Particularly at late times mergers of more massive galaxies are often ``dry'' with little amounts of cold gas involved, thus, hardly inducing high levels of nuclear activity.

Turning to moderately luminous AGN, the trends are somewhat different: from $z=2$ to $z=1$, the probability of hosting a moderately luminous AGN ($\sim$10~per cent) marginally decreases for galaxies without a recent merger event, but slightly increases for those with both major and minor mergers, suggesting that mergers may get more relevant for driving AGN activity in this redshift interval. Below $z=1$, the fractions of moderately luminous AGN are dropping down to 2~per cent at z=0.1 with mergers, and down to 0.5~per cent without mergers. The stronger decline of AGN fractions in galaxies without recent mergers points towards a slightly increased relevance of mergers for fuelling nuclear activity on a kpc-level in galaxies at and after $z=1$, although the probability that a galaxy with a recent merger event shows nuclear activity is still fairly low ($\lesssim$10~per cent).

Finally, we compare our simulation results with observed 
AGN fractions from \citet{Goulding_2018} and \citet{Juneau_2014}.
The black crosses show the datapoints from \citet{Goulding_2018} for major mergers and galaxies without any merger signatures (dashed and solid horizontal lines, respectively), including errorbars and the corresponding redshift range.
Since the mid-IR luminosities of the AGN sample in \citet{Goulding_2018} translate in bolometric luminosities between $3 \cdot 10^{43}$erg/s and $2 \cdot 10^{46}$erg/s with the majority being in the range $7 \cdot 10^{43}$erg/s - $5 \cdot 10^{45}$erg/s (private communication with A. Goulding), our simulated AGN fractions agree remarkably well with the observations.
The blue and yellow bars\footnote{The bars originate from measurements in different mass ranges and include all values for $M_*>10^{11} M_{\odot}$.} show fractions of local moderately luminous and luminous AGN, respectively,
obtained from an SDSS galaxy sample at low-redshift ($z \sim 0.1$) using optically-selected AGN from emission lines as described by \citet[][]{Juneau_2014}. The predicted AGN fractions of local galaxies are systematically lower by approximately half an order of magnitude. This rather minor difference might be caused by our limited resolution, also limiting the BH mass and thus the AGN luminosity. More likely, however, it is caused by selection effects, particularly since our AGN fractions agree remarkably well with the observations from \citet{Goulding_2018} and
since our AGN luminosity functions also agree very well with observations \citep{Hirschmann, Biffi_2018}.

%%%%%%%%%%%%%%%%%%%%%%%%%%%%%%%%%%%%%%%%
\subsubsection{The redshift evolution of merger fractions of AGN hosts}\label{mergerrates}
%%%%%%%%%%%%%%%%%%%%%%%%%%%%%%%%%%%%%%%

\begin{figure}
  \includegraphics[trim = 0mm 7mm 3mm 11mm, clip, width=0.47\textwidth]{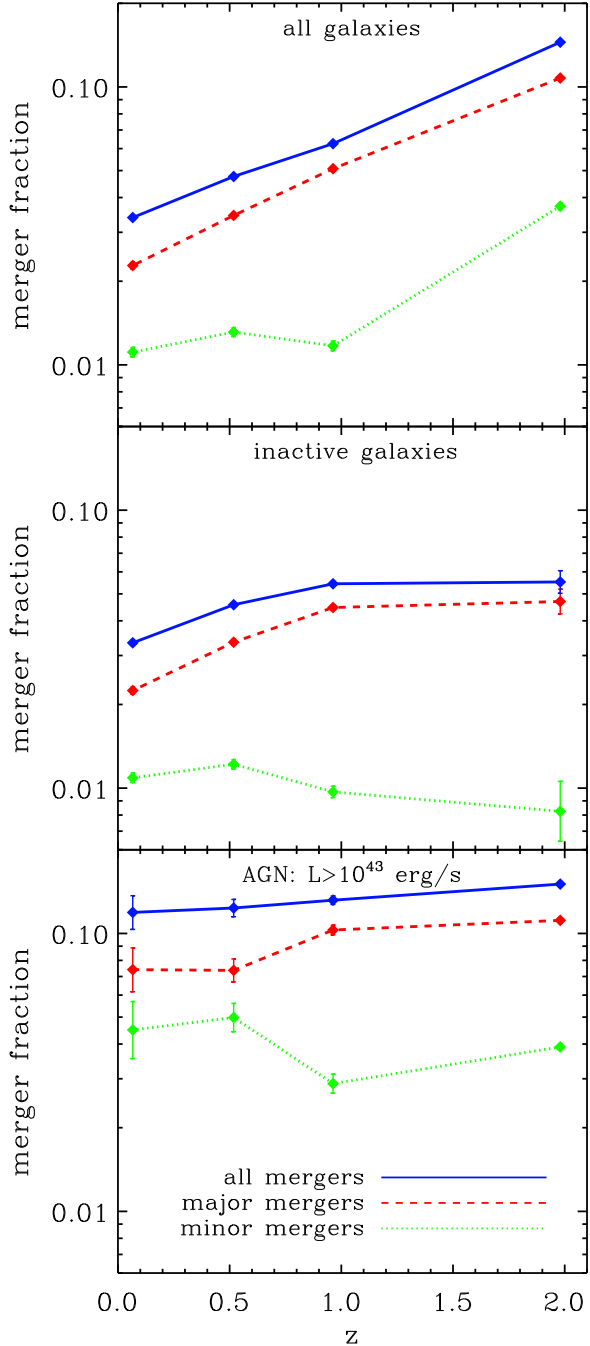}
  \caption{Redshift evolution of the total, major, and minor merger fractions (blue solid, red dashed, and green dotted lines, respectively) of all (top panel), inactive (middle panel), and active galaxies with $L_{\mathrm{bol}}>10^{43}$~erg/s (bottom panel), assuming a stellar mass cut of $M_*>10^{11} M_{\odot}$. Error bars indicate binomial confidence intervals.}
\label{mf_vs_z}
\end{figure}

After having demonstrated that at and below $z \sim 1$, mergers may induce nuclear activity in less than 10~per cent of galaxies (with recent mergers), in this subsection, we explore the probability that  AGN host galaxies have experienced a merger event in the past 0.5~Gyr, i.e. the total, minor, and major merger fraction of AGN hosts, $N_{\rm AGN,\ major+minor}/N_{\rm AGN}$, $N_{\rm AGN,\ major}/N_{\rm AGN}$ and, $N_{\rm AGN,\ minor}/N_{\rm AGN}$,\footnote{Note that the absolute value of the merger fraction strongly depends on the definition of mergers, i.e. during which time interval they are identified. We tested different time intervals of up to 1.5~Gyr, where the merger fraction is about twice as high as for our fiducial choice of 0.5~Gyr. Qualitative trends, however, remain unaffected.}. This quantity represents the {\it maximum} possible likelihood that the nuclear activity of an AGN population was fuelled (on a kpc level) by mergers. 

Fig. \ref{mf_vs_z} shows the redshift evolution of the total, major, and minor merger fractions (blue solid lines, red dashed lines, and green dotted lines, respectively) of AGN with $L>10^{43}$erg/s ($N_{\rm AGN, major+minor/major/minor}/N_{\rm AGN}$, bottom panel), compared to the merger fractions of inactive galaxies ($N_{\rm inactive,\  major+minor/major/minor}/N_{\rm inactive}$, middle panel) and to that of all galaxies, i.e. active {\it and} inactive ones ($N_{\rm AGN+inactive,\ major+minor/major/minor}/N_{\rm AGN+inactive}$, top panel).

For all galaxies, the total (major) merger fractions are strongly declining from  15 (10)~per cent at $z=2$ to less then 4 (3)~per cent at $z=0.1$. The predicted decrease of the total merger fractions from high to low redshifts is a direct consequence of an expanding, hierarchically growing Universe, and also qualitatively consistent with observations of \citet{Kartaltepe_2007} and \citet{Xu_2012} as well as with other simulation studies \citep[e.g., Millennium simulation,][]{Genel_2009}. Instead, the minor merger fractions hardly evolve with redshift, and stay always below 4~per cent at $z=0-2$. Such rather low minor merger fractions and their weak evolutionary trend may be caused by our definition of merger classes (galaxies in the major merger group can also have experienced minor mergers in the past 0.5~Gyr), not reflecting the {\em actual number} of major and minor mergers galaxies experienced during the past 0.5~Gyr.

When separating between active and inactive galaxies, total and major merger fractions of both active and inactive galaxies only exhibit a weak evolutionary trend, in contrast to all galaxies. In addition, active galaxies have on average a three times higher probability for a minor and/or major merger event in the recent past compared to inactive galaxies, whose total merger fractions stay always below 6~per cent. But also the merger fractions of active galaxies reach a maximum value of only 15~per cent, suggesting that the majority of nuclear activity of an AGN population at $z=0-2$ is unlikely to be caused by merger events.

%%%%%%%%%%%%%%%%%%%%%%%%%%%%%%%%%%%%%%%
\subsubsection{AGN merger fractions as a function of the AGN
  luminosity}\label{mergerrates_lum}
%%%%%%%%%%%%%%%%%%%%%%%%%%%%%%%%%%%%%%%

\begin{figure}  
  \includegraphics[trim = 0mm 8mm 0mm 17mm, clip, width=0.47\textwidth]{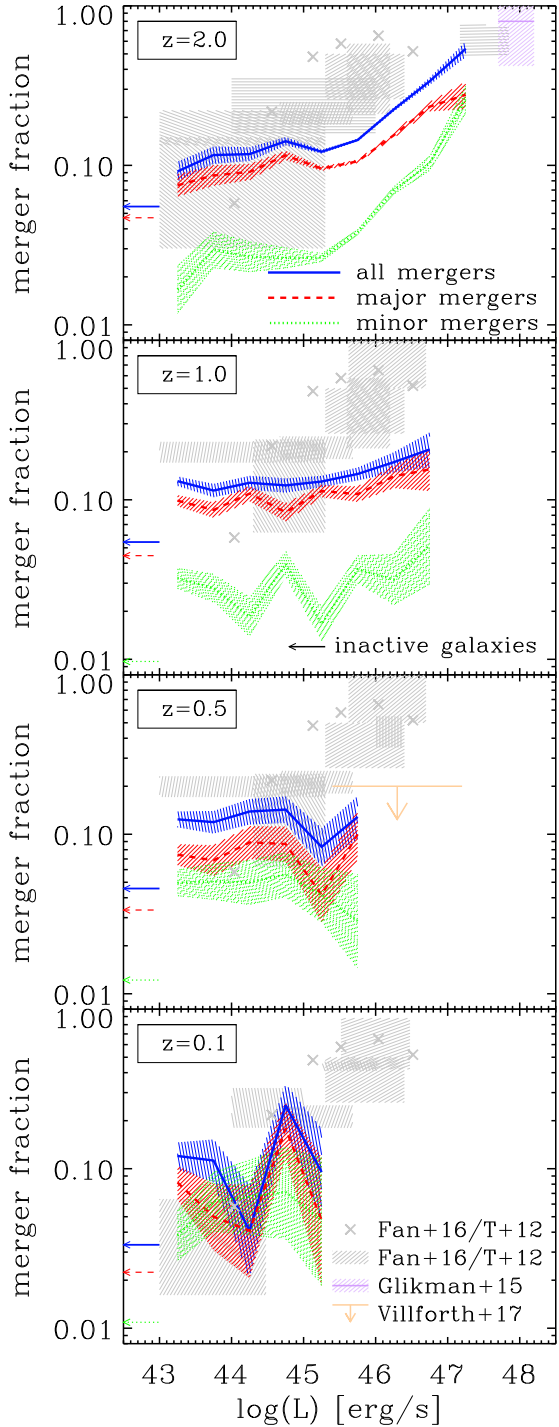}
  \caption{Total, major, and minor merger fractions and the corresponding binomial errors (blue solid, red dashed, and green dotted lines and shaded areas, respectively) of AGN host galaxies versus their bolometric AGN
           luminosity at different redshift steps ($z=2.0, 1.0, 0.5, 0.1$, panels from top to bottom) compared to that of inactive
           galaxies (depicted by the horizontal arrows at the
           left-hand side of each panel). Simulation predictions are
           compared to observed major merger fractions of AGN hosts
             (compilation of \citealt{Fan_2016}, including the datapoints from \citealt{Treister}: grey crosses and grey 
             shaded area, the latter illustrates the entire observed
             luminosity range and the error on the y-axis;
             \citealt{Glikman_2015}: purple horizontal line and shaded area; \citealt{Villforth_2016}: orange line
             with the arrow indicating the upper limit and the
             observed luminosity range).
	   }
\label{merger_fraction}
\end{figure}

To understand whether the maximum probability that an AGN was fuelled by a merger is related to the respective AGN luminosity, in Fig. \ref{merger_fraction}, we explore the total, major, and minor merger fractions as a function of AGN luminosity (blue solid, green dotted and red dashed lines, respectively) at different redshift steps ($z=2.0, 1.0, 0.5, 0.1$, panels from top to bottom). To avoid low number statistics, we consider only bins of AGN luminosity containing at least 20 active galaxies.
Fig. \ref{merger_fraction} shows that the global trends seen in Fig. \ref{mf_vs_z}, namely that total, major, and minor merger fractions of active galaxies are larger than that of inactive ones (illustrated by the arrows on the left-hand side in each panel of Fig. \ref{merger_fraction}), remain the same for each AGN luminosity, irrespective of the redshift step. 

Turning to the dependence of the merger fractions on AGN luminosity, at $z=2$ the total, major, and minor merger fractions strongly increase from less than 10, 8 and 2~per cent for faint AGN to up to more than 50, 30 and 30~per cent for most luminous AGN with $L_{\rm bol} \geq 10^{47}$~erg/s, respectively. Towards lower redshifts, at $z \leq 1$, the increase of the merger fractions with AGN luminosity is significantly weaker or even negligible, at maximum raising from 10~per cent for faint AGN up to 20~per cent for more luminous AGN. This trend may be
due to the fact that at and below $z=1$, even our large 500Mpc/hr simulation run does not contain sufficient statistics for AGN more luminous than $L_{\rm bol} \sim 5 \times 10^{46}$~erg/s at $z=1$, $L_{\rm bol} \sim 5 \times 10^{45}$~erg/s at $z=0.5$, and $L_{\rm bol} \sim 10^{45}$~erg/s at $z=0.1$, impeding us by construction to find
any potential increase of the merger fractions for these most luminous AGN. 

Compared to observed major merger fractions of the compilation of \citet[][grey crosses and grey shaded areas, illustrating the observed luminosity range and the uncertainty in the merger fraction]{Fan_2016}, including
observations from \citet{Treister} and \citet[][purple horizontal line and shaded area]{Glikman_2015}, we find at $z=2$ a qualitative (even if not quantitative) agreement between the  observed steep raise of the merger fraction towards higher AGN luminosities and our simulated AGN merger fractions.
In contrast, at lower redshifts ($z \leq 1$), the predicted dependence of the merger
fraction on AGN luminosity is significantly weaker than that
of \citet{Treister}, despite their rather large scatter at low
redshifts (due to low number statistics).
However, most of the observed data-points cover a very large redshift range, for example the grey crosses, making a comparison at specific redshifts difficult.
Furthermore, the different observed datapoints are the result of differently selected AGN samples. The compilation from \citet{Treister}, for example, consists of datapoints selected from X-ray, infrared (IR), and spectroscopic surveys. \citet{Fan_2016} add datapoints from dust-obscured, dust-reddened, and (mid-)IR-luminous quasars (see \citealt{Treister} and \citealt{Fan_2016} for details).
Compared to \citet[][orange line with the arrow indicating the upper limit of the merger fraction  and the observed luminosity range]{Villforth_2016}, our simulated major merger fractions of the 
most luminous AGN at $z=0.5$ are in good agreement with their maximum merger fraction of less than 20~per cent, being significantly lower than that of \citet{Treister} in the same luminosity range. We however emphasize that such a comparison between observed and simulated AGN merger fractions is complicated by a lot of caveats, not only due to the already mentioned various selection criteria, but also because of different merger identifications in observations and
simulations (see section \ref{obscomp} for further discussion).  

To summarize, except for very luminous AGN at $z=2$, our simulation predictions do not favour any prevalence ($>$50~per cent) of mergers for fuelling
nuclear activity in AGN populations at $z=0-2$, irrespective of the AGN luminosity. Nevertheless, the probability for AGN hosts of any AGN luminosity having experienced a major and/or minor event in the last 0.5~Gyr, can be up to three times higher than that for inactive galaxies. Such elevated merger fractions of active galaxies still point towards a connection between nuclear activity and merger events, even if mergers do not appear to be the statistically dominant fuelling mechanism for nuclear activity in our simulations.

%%%%%%%%%%%%%%%%%%%%%%%%%%%%%%%%%%%%%%%
\section{The dependence of AGN merger fractions on host galaxy properties}
\label{mergerrates_host}
%%%%%%%%%%%%%%%%%%%%%%%%%%%%%%%%%%%%%%%

In this section, we aim to understand the origin of (i) the slightly enhanced merger fractions of active galaxies, compared to that of inactive galaxies and (ii) the steep up-turn of AGN merger fractions towards high AGN luminosities at $z=2$, as shown in the last two sections \ref{mergerrates} and \ref{mergerrates_lum}. We explore to what extent these features of active galaxies can be explained by a combination of an intrinsic dependence of merger fractions on different galaxy properties, such as stellar mass and specific SFRs, and of a bias of AGN preferentially residing in galaxies with specific properties. To reveal that, we compare, {\em at fixed galaxy stellar mass or specific SFR}, the merger fractions of active to that of inactive galaxies, and we relate the former, the merger fraction of AGN, with the respective probability that  such AGN are hosted by galaxies of a given stellar mass or specific SFR.

\subsection{Galaxy stellar mass}\label{mstellar}
%%%%%%%%%%%%%%%%%%%%%%%%%%%%%%%%%%%%%%%

\begin{figure}
  \includegraphics[trim = 0mm 0mm 0mm 0mm, clip,
  width=0.5\textwidth]{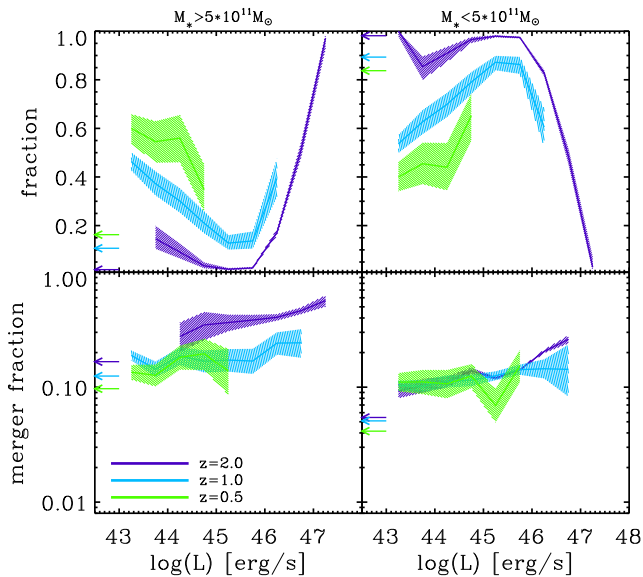}
  \caption{
	   Top row: Fraction of AGN residing in galaxies with $M_\star > 5 \times 10^{11} M_\odot$ and  $10^{11} M_\odot < M_\star < 5 \times 10^{11}
    M_\odot$ (top left and right panels, respectively) versus AGN luminosity at $z=0.5,1,2$ (differently coloured lines).  The shaded areas represent the corresponding binomial confidence intervals, and the arrows at the left-hand side of each panel indicated the fractions of inactive galaxies ($L_{\mathrm{bol}}<10^{43}$erg/s) in the two galaxy stellar mass bins. 
	   Bottom row: the same as in the top row, but for total AGN merger fractions (major {\em and} minor mergers) versus AGN luminosity at $z=0.5,1,2$ (differently coloured lines).}
\label{merger_fraction_mst}
\end{figure}

Starting with the dependence of AGN merger fractions on galaxy stellar mass, the bottom row in Fig. \ref{merger_fraction_mst} visualises the total AGN merger fractions (major and minor mergers) versus AGN luminosity at different redshift steps (differently coloured lines) separately for massive ($M_\star > 5
\times 10^{11} M_\odot$, left panel) and less massive host galaxies ($10^{11} M_\odot < M_\star < 5 \times
10^{11} M_\odot$, right panel). As seen for {\em all} galaxies/AGN in Fig. \ref{merger_fraction}, also for a given stellar mass bin, the merger fractions of AGN are elevated (by up to half a dex) at any redshift and AGN luminosity, compared to that of inactive galaxies (illustrated by arrows at the left-hand side of each panel). This implies that  at fixed galaxy mass (and thus, also at fixed BH mass), AGN hosts are also more likely to have experienced a recent merger than inactive galaxies, and thus, that nuclear activity of an AGN population can be  fuelled by merger events -- to a low degree, though, hardly exceeding 20~per cent. 

In addition, the bottom row in Fig. \ref{merger_fraction_mst} shows that AGN merger fractions of massive hosts are larger, by a factor of three at $z=2$, than that of less massive ones, at a given AGN luminosity and redshift. This difference is largely caused by the intrinsically up to half an order of magnitude higher merger fractions of  massive inactive galaxies compared to less massive ones (left-hand arrows). This dependence of merger fractions on the galaxy stellar mass is a natural consequence of a hierarchically growing Universe, in which massive galaxies experience a much more complex merger history than low mass galaxies \citep[e.g.][]{Fakhouri_Ma_2008, Genel_2009}.

Interestingly, at a given host stellar mass the AGN merger fraction is at any redshift largely independent of the AGN luminosity. At $z=2$, this is in stark contrast to the strongly raising  merger fractions of {\it all} AGN hosts towards higher AGN luminosity, as shown in the top panel of Fig.  \ref{merger_fraction}. To understand this difference, we have to consider the probability that an AGN resides in a massive or less massive host as a function of the AGN luminosity (see top row of Fig. \ref{merger_fraction_mst}). While most luminous AGN (with $L_{\rm bol} > 3 \times 10^{46}$~erg/s) are preferentially hosted by massive galaxies at $z=2$, less luminous AGN are mostly living in less massive galaxies (see lilac curves in top panels of Fig. \ref{merger_fraction_mst}). Thus, this bias in AGN host stellar mass, together with the intrinsic dependence of merger fractions on the galaxy stellar mass, can, to some extent, explain the steep up-turn of AGN merger fractions towards higher AGN luminosities at $z=2$. In other words, the high merger fractions of luminous AGN at $z=2$ partly reflect the intrinsically higher merger fractions of massive galaxies, in which luminous AGN predominantly reside. Note that this result is consistent with recent findings from phenomenological models of \citet{Weigel_2018}. Nevertheless, as pointed out before, the more than twice as large merger fractions of luminous AGN at $z=2$ (ca 50~per cent) compared to that of massive inactive galaxies (ca 20~per cent), still indicate the relevance of mergers for fuelling nuclear activity in most luminous AGN.

\subsection{Specific star formation rate}\label{ssfr}
%%%%%%%%%%%%%%%%%%%%%%%%%%%%%%%%%%%%%%%

\begin{figure}
\includegraphics[trim = 0mm 0mm 0mm 0mm, clip,
width=0.5\textwidth]{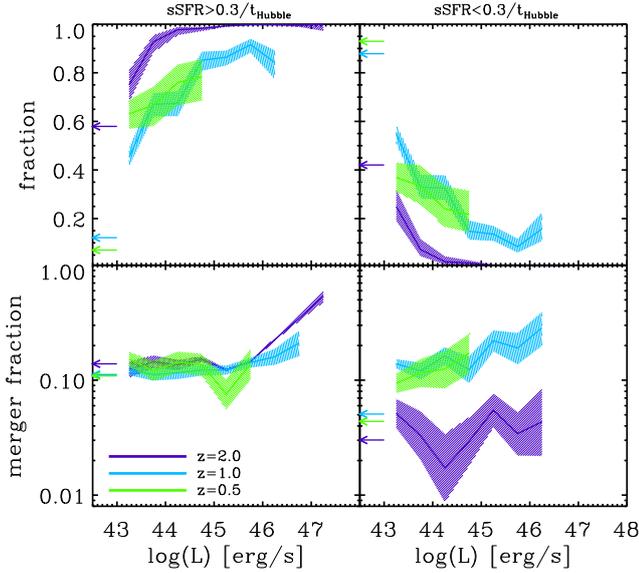} 
  \caption{Same as Fig. \ref{merger_fraction_mst}, but when
    distinguishing between star-forming (left column) and quiescent
    galaxies (right column) with specific SFR
    $ > 0.3/t_{\mathrm{Hubble}}$ and specific SFR
    $ < 0.3/t_{\mathrm{Hubble}}$, respectively.
	  }
\label{merger_fraction_ssfr}
\end{figure}

Next, we turn to the dependence of AGN merger fractions on the specific SFRs of their hosts, i.e. to what extent AGN merger fractions are different for star-forming (SF) and passive galaxies, i.e. galaxies with specific SFRs above and below $0.3/t_{\mathrm{Hubble}}$, respectively.

The bottom left panel of Fig. \ref{merger_fraction_ssfr} shows that the AGN merger fractions of SF hosts at $z=0.5, 1, 2$ (differently coloured lines) are widely independent of AGN luminosity,  except for the up-turn of the merger fractions for the most luminous AGN at $z=2$,\footnote{This up-turn is a consequence of luminous galaxies being mostly hosted by {\em massive} SF galaxies (see Fig. \ref{merger_fraction_mst}).} and have very similar values ($10-20$~per cent) as the merger fractions of inactive SF galaxies (arrows on the left). Moreover, as the top left panel of Fig. \ref{merger_fraction_ssfr} illustrates, AGN predominantly reside in SF galaxies, in particular at $z=2$ ($>$80~per cent) and to lesser extent also at $z=1$ ($>$70~per cent) and $z=0.5$ ($>$60~per cent). These results suggest that star formation and nuclear activity are related on a statistical level, and both SF/starbursts, and BH fuelling may be induced by merger events (on average up to 10-20~per cent of AGN/SF galaxies). The generally higher merger fractions of {\em all} active compared to {\em all} inactive galaxies, i.e. not distinguishing between SF and passive galaxies (see e.g., Fig. \ref{mf_vs_z}), thus reflect the intrinsically higher merger fractions of SF galaxies, in which AGN predominantly occur.
This is largely consistent with observations finding a close link between AGN activity and star formation activity \citep[e.g.][]{Juneau_2013}.
  
AGN merger fractions of passive hosts are half as high as that of SF hosts at $z=2$, while at $z \leq 1$ they are similar to that of SF hosts. In addition, for passive galaxies, AGN merger fractions are always higher (by ca 0.5dex) than the merger fraction of inactive galaxies, suggesting that a
merger may raise the gas supply and density within the central few
kpc, but the gas does not get cold or dense enough to induce significant
levels of SF. Note that per se nuclear activity in passive galaxies can be
explained by (i) warm/hot gas being accreted on the central BH, not fullfilling
SF criteria, and (ii) the computed Bondi accretion
rate's strong dependence on BH mass so that for massive BHs, already
small amounts of gas and lower gas densities are sufficient to ignite
 moderately luminous AGN. However, only a small fraction ($<$ 30~per cent) of passive galaxies host moderately luminous AGN, and less than 10~per cent of passive galaxies host luminous AGN, showing that it is not very likely to have nuclear activity in galaxies without on-going SF.

To summarize section \ref{mergerrates_host}, the high merger fractions of luminous AGN at $z=2$ in the top panel of Fig. \ref{merger_fraction}, reflect, on the one hand, the intrinsically high merger fractions of massive galaxies, and on the other hand, an enhanced role of mergers for providing the gas fuel in the central few kpc for BH accretion. The generally elevated merger fractions of active with respect to inactive galaxies (Figs. \ref{merger_fraction} and \ref{mf_vs_z}) are to a large degree connected to the intrinsically high merger fractions of SF galaxies, in which AGN primarily appear. Also at any given galaxy stellar mass or specific SFR, higher merger fractions of active galaxies (but on average not exceeding 20~per cent, except for luminous AGN at $z=2$), compared to inactive passive galaxies, indicate only a weak, albeit still non-negligible role of mergers for nuclear activity (and star formation).

%%%%%%%%%%%%%%%%%%%%%%%%%%%%%%%%%%%%%%%
\section{Central gas properties and BH masses in (in)active galaxies with different merger histories}
%%%%%%%%%%%%%%%%%%%%%%%%%%%%%%%%%%%%%%%
\label{gas_properties}

\begin{figure*}
  \includegraphics[trim = 0mm 5mm 0mm 0mm, clip, width=0.9\textwidth]{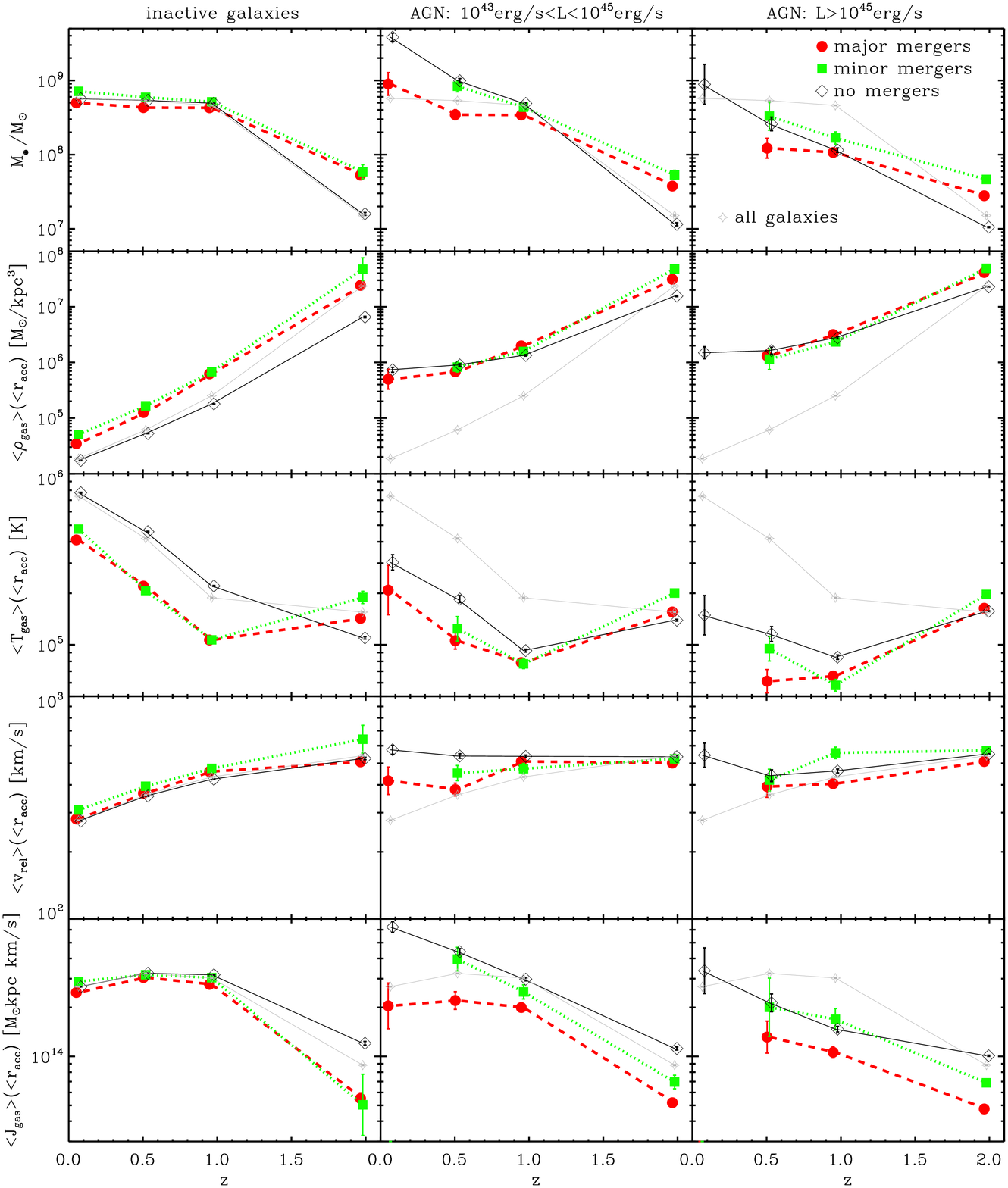}
  \caption{Redshift evolution of the mean of the BH mass,  density,  temperature, relative  velocity and  angular momentum of the gas within the resolved accretion regions around the BH (rows from top to bottom) for inactive galaxies (left panels), moderately luminous AGN (middle panels), and high-luminosity AGN (right panels), having experienced either a recent major merger (filled red circles), minor merger (filled green squares), or no merger (black open diamonds). 
For comparison, the grey diamonds and solid lines show the mean values for the total galaxy sample, irrespective of AGN activity and merger history (thus, they are the same for each row).
All parameters are computed at the time of the snapshot, when the merger has been identified, or in case of "no mergers", 0.5~Gyr before the respective redshift.
  Error bars indicate the bootstrapping errors. For better readability,  symbols and error-bars are slightly shifted around the redshift-values $z=0.1$, $z=0.5$, $z=1.0$, and $z=2.0$. 
	  }
\label{meanparms_Bondi}
\end{figure*}

Up to now, we have shown that the fraction of active galaxies having recently experienced a merger event is generally larger than that of inactive galaxies, indicating that mergers may fuel nuclear activity on a kpc-level. In this section, our goal is to obtain a deeper physical understanding for this result, by investigating the quantities governing the accretion rates onto BHs in our simulations, i.e. used to compute the Bondi accretion rate by virtue of eq. \ref{accretion_rate}: the BH mass, the gas density, the gas temperature, and the gas velocity relative to the BH {\em within the resolved accretion region}, $r_{\mathrm{acc}}$\footnote{The accretion radius (not to confuse with the Bondi radius) is defined as the radius, inside which the gas particles are used to compute the Bondi accretion rate. Since the number of gas particles used for that calculation is fixed, the accretion radius varies for different BHs.}. Specifically we  address the following two questions:
\begin{enumerate}
\item Which central ISM conditions around the BHs and which BH masses in our modelling approach are necessary for causing nuclear activity in galaxies, i.e. how do ISM conditions and BH masses differ between active and inactive galaxies?
\item Which ISM pre-conditions are necessary such that mergers may lead to nuclear activity in galaxies, i.e. in what way do central ISM conditions and BH masses of merging active galaxies differ from non-merging active galaxies?
\end{enumerate}
To robustly address these two questions, in particular (ii),  we consider the central gas properties and BH masses shortly (at the snapshot) {\em before} the merger happened or, for galaxies without a recent merger, shortly before the considered redshift\footnote{For $z \ge 1$, the typical times to the previous snapshot are $\sim 0.5$~Gyr, and for $z < 1$, they range between 0.3 and $0.4$~Gyr. Note that the exact time interval varies slightly, because it depends on the next time step with synchronous output.}. This allows us to disentangle the effect of the general underlying gas properties of the host galaxies from the effect of the mergers, which can strongly influence these gas properties (increasing gas density, reducing gas temperature and relative velocity).

Naively, we would expect that the ISM properties would scale {\em by construction} with the accretion rate onto the BH and thus with AGN activity. However, as we shall see, the complex interplay between various physical processes in cosmological simulations, such as AGN and stellar feedback, gas cooling and the related in-flowing cold gas streams, disproves such an expectation.

Fig. \ref{meanparms_Bondi} shows the redshift evolution of the mean BH mass, gas density, gas temperature, relative gas velocity and angular momentum of the gas within $r_{\mathrm{acc}}$ (rows from top to bottom), separating between inactive galaxies, moderately luminous, and luminous AGN (left, middle, and right columns, respectively), to address question (i). To also investigate point (ii), we additionally split the samples into galaxies/AGN hosts with major (red filled circles), minor (green filled squares), and no merger events (black open diamonds).
For better comparison, we also show the mean values for the {\em total} galaxy sample (grey diamonds and solid lines) in each panel, irrespective of AGN activity and merger history. Unsurprisingly, these average quantities of all galaxies most closely resemble that of non-merging active galaxies at $z=2$ (black lines in middle and right columns) and that of non-merging inactive galaxies at $z \leq 1$ (black line in left column).

The first row of Fig. \ref{meanparms_Bondi} shows that at $z=2$, the average BH masses are not significantly different for active and inactive galaxies, at a given merger class. Towards lower redshift, at $z=1$ and $z=0.5$, the situation changes: more luminous AGN host on average {\em less} massive BHs than moderately luminous AGN and inactive/all galaxies, {\em irrespective of the merger history}. Thus, a higher AGN luminosity is improbably caused by a larger BH mass. Moderately luminous AGN without any merger at low redshifts, in particular at $z=0.1$, have by a factor of three higher BH masses than inactive galaxies without any merger or all galaxies, indicating that large BH masses in galaxies without any mergers can promote nuclear activity at moderate levels.

Turning to the gas density within $r_{\mathrm{acc}}$, the second row in Fig. \ref{meanparms_Bondi} illustrates that this quantity generally decreases from high to low redshifts, for all, merging, and non-merging galaxies/AGN. Contrasting the gas densities of inactive/all with that of active galaxies, at $z=2$, we find hardly any difference, in particular for galaxies, which will be under-going a merger, where the mean inner gas density is always larger than $10^7 M_{\odot}$/kpc$^3$. The generally high central gas densities in galaxies at $z=2$ favour AGN activity independently of merger events, leading to the high AGN fraction shown in Fig. \ref{AGN_fraction_z}. A small fraction of galaxies, though, do not reach the threshold for being an AGN ($L>10^{43}$erg/s) despite the high inner gas densities shortly before the merger. Towards lower redshifts $z \leq 1$, the central gas densities of active galaxies stay on average, irrespective of the merger history, at or above  $10^6 M_{\odot}$/kpc$^3$, and they are by more than one order of magnitude higher than that of inactive or all galaxies, which, instead, drop below $10^5 M_{\odot}$/kpc$^3$ towards $z=0$. This demonstrates that an enhanced gas density is a necessary (but not sufficient) pre-condition for nuclear activity, even if a galaxy will experience a merger event. 

Comparing the gas densities of merging and non-merging galaxies, we find that merging, inactive galaxies have  by a factor of 5 increased central gas densities compared to non-merging inactive galaxies. Interestingly, active galaxies instead, shortly before having a major or minor merger, have similarly high gas densities as those without any merger event, suggesting that {\it central gas densities can be sufficiently increased not only by merger events, but also by other processes} (see discussion \ref{discussion}).

Exploring the mean gas temperatures within $r_{\mathrm{acc}}$ (third row of Fig. \ref{meanparms_Bondi}) largely reveals opposite trends compared to the gas densities: the gas temperatures  {\em increase} towards $z=0$, as gas gets heated by various heating processes (e.g. gravitational heating, AGN feedback), simultaneously becoming less and less dense. On average and irrespective of the presence of a merger, active galaxies have lower ($<3\times 10^5$~K) inner gas temperatures than inactive or all galaxies, at least at $z \leq 1$, resulting in higher BH accretion rates (see eq. \ref{accretion_rate}).
Still for $z \leq 1$, the average gas temperature right before a merger is reduced in both active (at maximum $2\times 10^5$~K) and inactive galaxies (at maximum $5\times 10^5$~K) compared to {\em non-merging} active/inactive galaxies, possibly as a consequence of (pre-)merger-induced cooling flows.
 
Considering the relative gas velocities $v_{\mathrm{rel}}$ within $r_{\mathrm{acc}}$ (fourth row of Fig. \ref{meanparms_Bondi}), at $z \leq 1$, this quantity is by a factor of up to 3 higher for active galaxies, at least when they have no merger or only a minor merger, than for inactive or all galaxies. This is surprising as, by construction, a higher relative gas velocity {\em decreases} the Bondi accretion rate (eq. \ref{accretion_rate}). A high relative gas velocity may, however, indicate increased gas inflows towards the centre. Such inflowing gas does not only seem to counteract the intrinsically reduced Bondi accretion rate, but also appears to be crucial to provide sufficient fuel to induce nuclear activity (in galaxies with mergers as well as without mergers).

Tightly connected to the relative gas velocity is the angular momentum of the gas (bottom row in Fig. \ref{meanparms_Bondi}), even if not explicitly considered, when estimating the Bondi accretion rate. While at $z=2$ the mean angular momentum is not significantly different in active and inactive galaxies, at $z \leq 1$ the mean angular momentum of gas in luminous AGN hosts is lower than that in moderately luminous AGN hosts and inactive galaxies, showing that a lower angular momentum of the gas promotes strong nuclear activity.
Over the entire redshift range, active galaxies right before a major merger (and to lesser extent, also before a minor merger) have a by up to a factor of three lower angular momentum of the central gas than active galaxies {\em without} a merger, possibly due to the (on average) different environments of merging and non-merging galaxies.

To summarise this section, to induce significant levels of AGN activity in galaxies, comparably high central gas densities, and low gas temperatures are prerequisites. Since at $z \leq 1$, these ISM properties already differ on average significantly right before the merger between active and inactive galaxies, nuclear activity in merging galaxies is not necessarily related to the merger event.
Compared to non-merging AGN hosts, active galaxies under-going a (major) merger are largely characterised by having lower gas temperatures and lower relative velocities, possibly due to (pre-)merger-induced cooling flows, promoting nuclear activity. Instead, the higher gas temperatures and higher relative velocities of non-merging AGN hosts, in particular for moderately luminous AGN, are likely compensated by higher BH masses, resulting in similar levels of nuclear activity as for merging AGN hosts.

%%%%%%%%%%%%%%%%%%%%%%%%%%%%%%%%%%%%%%%
\section{Discussion}
%%%%%%%%%%%%%%%%%%%%%%%%%%%%%%%%%%%%%%%
\label{discussion}

In this section, we discuss our results with respect to (i) the importance of mechanisms other than mergers for driving nuclear activity (section \ref{environment}), (ii) limitations and caveats of our analysis (section \ref{caveats}),  and (iii) to what extent our results (dis)agree with observations (\ref{obscomp}) and with previous model predictions (semi-analytic and semi-empirical models, section \ref{modelcomp}).

\subsection{AGN fuelling processes: the role of the large-scale environment}\label{environment}

Since our simulation predictions indicate that more than $\sim$80~per cent of AGN, in host galaxies with stellar masses above $M_* > 10^{11} M_{\odot}$
cannot be fuelled by mergers (except for AGN more luminous than $10^{46}$~erg/s at $z=2$), the question arises which mechanisms instead predominantly cause nuclear activity. In cosmological simulations, AGN activity can be principally driven by smooth  accretion of gas originating from cooling from a hot halo, from mass loss via stellar winds, or gas inflows from and, thus, depending on the large-scale filamentary structure\footnote{Note that gas flows via violently unstable disks and/or secular evolution disk instabilities cannot be resolved in our simulations.}. While a detailed analysis of the relative importance of such different mechanisms clearly goes beyond the scope of this paper, we briefly discuss the possible importance of the environment/filamentary structure of galaxies on their nuclear activity.

When employing an often used density criterion to characterise the environment (number counts of neighbouring galaxies within 1 or 2 Mpc), we do not find any relation between the central gas density (governing BH accretion) and the density of the environment. Instead, \citet{Steinborn_2016} showed that to specifically study the role of the filamentary structure, the environment is well characterised by "tracing back" gas inflows: \citet{Steinborn_2016} analyse 34 dual AGN, offset AGN and inactive BH pairs at $z=2$ extracted from the Magneticum Pathfinder Simulations. They find that dual AGN on average accrete more gas originating from the surrounding medium (e.g. from filaments) than offset AGN or inactive BH pairs, suggesting the AGN activity is indeed correlated to "external"  gas accretion (opposed to stellar mass loss and halo gas cooling) from large-scale filaments. To robustly address this issue, we plan to relate the gas density at large radii to that in the inner region in future work.

\subsection{Caveats of large-scale cosmological simulations}\label{caveats}

All state-of-the-art cosmological simulations, including the Magneticum simulations considered in this work, generally suffer from limited resolution ($>0.7$~kpc) and adopt phenomenologically motivated sub-grid schemes to model small-scale physical processes, such as BH accretion and AGN feedback. Here we discuss potential caveats originating from these short-comings for our analysis.
  
\subsubsection{Inner gas flows and BH accretion}

Due to limited resolution in cosmological simulations, innermost gas inflows ($<$ kpc) onto the central BHs cannot be resolved, likely affecting the resulting AGN luminosities, and potentially causing some further delay between the merger event and the peak in BH accretion. We additionally cannot resolve inner gas flows due to violently unstable discs, or secular evolution disk instabilities, impeding us to draw any conclusion on their potential role for causing AGN activity. 

BH accretion is estimated by the idealized Bondi model by virtue of equation (\ref{accretion_rate}), which is known to be a good approximation just for spherical accretion (i.e., for hydrostatic hot gas).
However, in addition to cosmological simulations hardly resolving the Bondi radius,
the Bondi scheme seems to also be a poor model for describing the accretion of cold, turbulent gas \citep[e.g.,][]{Hopkins_Quataert_2011, Gaspari, Steinborn_2015, Angles_Alcazar_2017}. Thus, particularly at higher redshifts and/or lower mass galaxies, when a lot of cold gas is likely to be accreted onto the BH \citep{Hopkins_Quataert_2011}, AGN luminosities could strongly be affected. Also increasing the resolution, which decreases the accretion radius, can influence BH accretion rates and AGN luminosities due to changing gas properties in the vicinity of the BH.
Even if adopting different BH accretion models or increasing the resolution would not affect merger histories of AGN hosts, AGN merger fractions could change, because of the dependence of the AGN luminosities on the accretion model/resolution. Nonetheless, we do not expect that such modifications would dramatically increase AGN merger fractions so that merger events would still play only a minor role for fuelling nuclear activity.

\subsubsection{AGN feedback}

To model AGN feedback, a fraction of the released accretion energy is injected into the ambient medium as a purely thermal energy input. \citet{Steinborn_2015} and \citet{Hirschmann} have shown that such an AGN feedback scheme is a bit too inefficient, resulting in too many too massive and too star-forming galaxies. Moreover, even if the evolution of AGN luminosity functions is well reproduced \citep{Hirschmann}, we over-estimate the number density of massive, radiatively efficient BHs at low z \citep{Schulze_2015}. A different AGN feedback model, which regulates more efficiently the gas content around massive BHs in massive galaxies \citep[see, e.g.,][]{Weinberger_2017, Choi_2017} would lead to an earlier shut-down of AGN. As a result, at low redshifts, the amount of AGN originating from smooth gas accretion onto a massive BH might be reduced, which may slightly increase AGN merger fractions. To test this hypothesis, for the future, we plan to run a new simulation set with an improved AGN feedback model, where the effect of the feedback model on the AGN merger fractions can be investigated in detail.

\subsection{Comparison to observations}\label{obscomp}

We have demonstrated that the predictions from our simulations are consistent with recent observations, in the sense that the majority of nuclear activity is unlikely caused by merger events. Simulations can also reproduce the observed increase of AGN merger fractions with increasing luminosity (\citealp{Treister}, \citealp{Fan_2016}) at $z=2$, but not at $z \leq 1$. These observations are, however, collected from different data sets of various studies in different redshift and luminosity ranges, applying different selection criteria
\citep{Bahcall_1997, Urrutia_2008, Georgakakis_2009, Koss_2010,
  Kartaltepe_2010, Cisternas_2011, Schawinski_2011, Kocevski_2012,
  Schawinski_2012, Lanzuisi_2015, Kocevski_2015, Hong_2015,
  Glikman_2015, DelMoro_2016, Wylezalek_2016}.
Thus, a quantitative comparison of merger fractions in simulations and these observations is complicated by two main reasons: (i) the huge variety of different selection criteria adopted in various observational studies and (ii) the intrinsically different merger identifications in observations and simulations. 

Regarding the latter, we adopt a specific definition for tagging a galaxy to be a major or minor merger in simulations: the \textsc{subfind} algorithm defines the exact snapshot/time, at which two galaxies are bound to each other for the first time, such that  the exact merger history of AGN hosts can be quantified. How long galaxies/AGN hosts are traced back in time to identify mergers, i.e. 0.5~Gyr, is a choice we made to capture typical timescales of galaxy mergers.

In profound contrast, in observations the identification of merger events is usually done on a visual basis at the same time the AGN luminosity is measured, thus neglecting any potential delay between the merger and significant levels of nuclear activity. A further consequence of a visual merger classification is that mostly/only major mergers can be identified, since minor mergers are not resolved properly and/or do not leave any clear visual signature in the morphological structure of a galaxy. These limitations of observations imply that observed merger detections might be underestimated, compared to our theoretical definition in simulations. For an accurate comparison between simulations and observations, a construction of mock images would be necessary, applying the same visual merger classification criteria and combining them with other observational selection criteria -- clearly beyond the scope of this study.

\subsection{Comparison to previous theoretical predictions}\label{modelcomp}

In previous studies, both semi-empirical as well as semi-analytic models have been used to investigate the relevance of different fuelling mechanisms, including merger events, for nuclear activity in galaxies. We now briefly discuss, how previous results compare to our findings in this work.

\subsubsection{Semi-empirical models}
The very first tools to study BH evolution in a statistical context have been phenomenological and semi-empirical models. These are characterized by a bottom-up approach. The least possible assumptions and associated parameters initially define the models. Gradually, additional degrees of complexities can be included, wherever needed.
In semi-empirical models \citep[e.g.][]{Hopkins_2009, Zavala_2012, Shankar_2014} galaxies (and eventually their central BHs) are not grown from first principles but they are assigned to host dark matter haloes via abundance matching techniques \citep[e.g.][]{Vale_Ostriker_2004, Shankar_2006} and allowed to merge following their dark matter merger trees. 

Among the results obtained from these type of models more relevant to the present work, we recall: (i) the declining AGN duty cycle and characteristic Eddington ratio of active BHs with time, possibly following an overall cosmic starvation \citep[e.g.][]{Shankar_2013}; (ii) the relatively minor role of mergers in building galaxies (and their BHs) with stellar mass $\log M_{\mathrm{stellar}} \leq 11 M_\odot$ (e.g., \citealt{Lapi_2018} and references therein); (iii) the key role of AGN feedback in shaping in particular the most massive galaxies \citep[e.g.][]{Fiore_2017}. 

Semi-empirical models have shown that galaxy-galaxy mergers can easily account for the vast majority of AGN at least at $z>1$ \citep[e.g.][]{Wyithe_2003, Shen_2009}. However, at high redshifts and high masses, haloes are rarely destroyed once formed \citep[e.g.][]{Sasaki_1994}. Thus halo merger fractions can be also viewed more straightforwardly as halo formation rates, usually conducive to gas-rich and rapid galaxy/BH formation episodes \citep[e.g.][]{Granato_2004, Lapi_2006, DiMatteo_2012}. Only at $z<1-1.5$ the merger/halo formation model starts breaking down and becoming distinct from more general gas-rich galaxy/BH triggering events \citep[e.g.][]{Menci_2003, Vittorini_2005, Draper_Ballantyne_2012}. Thus, all semi-empirical studies tend to align with the conclusion that intermediate-to-major mergers may fall short in accounting for the full statistics of low-luminosity AGN at $z<1$ \citep[e.g.][]{Scannapieco_Oh_2004, Shen_2009, Draper_Ballantyne_2012}.

\subsubsection{Semi-analytic models}
In contrast to phenomenological and semi-empirical models, SAMs populate dark matter halos, following their merger histories, with galaxies and BHs via modelling baryonic processes {\em from first principles}.  Historically motivated by binary merger simulations, "last-generation", but also most "state-of-the-art" SAMs \citep{Somerville, Croton06, DeLucia_2007, Bonoli_2009, Henriques_2015, Hirschmann16} assume that AGN activity is purely triggered by merger events \citep[see, however,][]{Bower06}, even though different implementations regarding minor/major mergers and BH growth curves have been developed. 
Such merger-driven BH models disagree with the results from cosmological simulations, presented in this work. 

It has been, however, repeatedly shown that adopting a purely merger-driven BH growth scenario in SAMs largely fails to reproduce the evolution of the observed AGN luminosity function and the corresponding antihierarchical trend in BH growth, due to severely underestimating the number density of faint/moderately luminous AGN at low redshifts \citep[see, however,][]{Bonoli_2009}. To overcome this deficiency, different solutions have been proposed: nuclear activity has been adopted to be additionally driven by (i) secular evolution disk instabilities \citep{Hirschmann_2012}, (ii) galaxy fly-bys \citep{Menci_2012}, and/or (iii) hot gas accretion onto the BH (ADAF model, \citealt{Fanidakis12}), or a combination of these processes. In most of these enhanced SAMs, merger events are, however, still necessary to predict a large enough amount of most luminous AGN \citep{Hirschmann_2012, Menci_2014} -- a trend, qualitatively consistent with cosmological simulations (at least at $z=2$). Overall, in SAMs (as in cosmological simulations), it remains unclear, which is the main driving mechanism for the majority of (moderately luminous) AGN.

\section{Conclusion}
\label{conclusion}
In this work, we theoretically investigated the statistical significance of merger events for fuelling nuclear activity (on scales of a few kpc) in galaxies at  $z=0-2$. To conduct this analysis, we employed two cosmological hydrodynamic simulations from the Magneticum Pathfinder Simulation set: first, a simulation with a comparably small volume of $(68 \mathrm{Mpc})^3$, but a resolution high enough to resolve galaxies' morphological structures, was used to explore the light curves of the central BHs of six individual example galaxies. Secondly, another simulation run, featuring large populations of even the most luminous AGN, thanks to a fairly large cosmological volume of $(500 \mathrm{Mpc})^3$, allowed us to study the relevance of mergers for fuelling nuclear activity over a wide AGN luminosity range in a global statistical context.

Analyzing our five test cases showed that merger events may significantly increase the probability for nuclear activity of a galaxy, but they do not necessarily boost the accretion onto BHs. In fact, analyzing the effect of a merger on nuclear activity is complicated by the high time-variability of BH accretion/AGN luminosity. To still perform a meaningful analysis, we investigated the effect of the recent merger history on AGN luminosity {\em for a statistically large sample of AGN} at a given time-step. Specifically, we can summarise the following main results:
\begin{itemize}

\item In galaxy populations, recent major/minor events can increase the probability for nuclear activity in galaxies by up to half an order of magnitude at $\leq$ 1, never exceeding 20~per cent though, compared to that of galaxies with a quiet accretion history. At $z \sim 2$, instead, irrespective of the merger history, almost all galaxies contain an AGN,  thanks to large amounts of dense gas present in galaxies at these early epochs.

\item In AGN populations, {\em mergers cannot be the statistically prevalent fuelling mechanism for  nuclear activity at $z = 0 - 2$ (hardly ever exceeding 20~per cent)}, except for very luminous AGN ($L_{\rm bol} > 10^{46} erg/s$) at $z \sim 2$. The high merger fractions ($>$ 50~per cent) of such very luminous AGN at $z = 2$ reflect, however, to some extent intrinsically high merger fractions of massive galaxies, in which luminous AGN preferentially reside.

\item Despite the statistically minor relevance of mergers for nuclear activity, the probability for AGN hosts to have experienced a recent major and/or minor merger event can be {\em by up to three times higher} than that for inactive galaxies. Such elevated merger fractions of active galaxies still point towards a connection between nuclear activity and merger events -- consistent with the expectations from binary merger simulations.

\item Investigating the ISM properties (gas density, gas temperature, relative velocity between BH and gas) in the vicinity of BHs shows that comparably high central gas densities and low gas temperatures are required (partly by construction via equation \ref{accretion_rate}) to induce nuclear activity in galaxies. Such prerequisites can be already present right before a merger and thus, they are not necessarily caused by a merger event. 

\item Active, merging galaxies are characterised by lower gas temperatures and relative velocities compared to active non-merging galaxies, promoting nuclear activity. The higher gas temperatures and relative velocities of non-merging AGN hosts, instead, are compensated by higher BH masses, still enabling nuclear activity at moderate luminosities.

\end{itemize}

We conclude that, even if mergers may increase the probability for nuclear activity by a factor of three, they still play only a minor role for causing nuclear activity in the overall AGN population ($<$ 20~per cent). This result is in profound disagreement with the traditional theoretical view, favouring a predominantly merger-driven BH growth/AGN activity, but it is consistent with a number of recent observational studies. 

Despite this progress, our simulations/analysis do not allow us to draw any robust conclusion on the dominant fuelling mechanisms for AGN activity (disk instabilities, smooth accretion from hot halo, cold inflows, stellar mass loss etc.) and on the processes, which are actually driving the gas onto the central BHs at sub-kpc and sub-parsec scales, because of limited resolution and  phenomenologically motivated models for BH accretion and feedback. Future theoretical studies performing "precision" cosmological simulations, by unifying a cosmological framework with the accuracy of detailed, small-scale simulations for modelling BH accretion and AGN feedback, will be certainly necessary to obtain a full understanding of the relative, statistical importance of different fuelling/triggering mechanisms for nuclear activity.

\section*{Acknowledgements}
We thank Andreas Burkert, Andy D. Goulding, and Sara L. Ellison for fruitful discussions during the preparation of this paper.

This research was supported by the DFG Cluster of Excellence `Origin and structure of the universe' and the SFB-Tansregio TR33 `The Dark Universe'.
MH acknowledges financial support from the European Research Council via an Advanced Grant under grant agreement no. 321323?~@~Z NEOGAL.
MK acknowledges support by DFG grant KR 3338/3-1.

We are especially grateful for the support by M. Petkova through the Computational Center for Particle and Astrophysics (C2PAP). Computations have been performed at the at the ’Leibniz-Rechenzentrum’ with CPU time assigned to the Project ’pr86re’ as well as at the ’Rechenzentrum der Max-Planck-Gesellschaft’ at the ’Max-Planck-Institut fu\"r Plasmaphysik’ with CPU time assigned to the ’Max-Planck-Institut fu\"r Astrophysik’.

\bibliography{AGN_triggering}

\section*{Appendix: estimation of the merger mass ratio}
\begin{figure*}
  \includegraphics[trim = 1mm 10mm 1mm 10mm, clip, width=\textwidth]{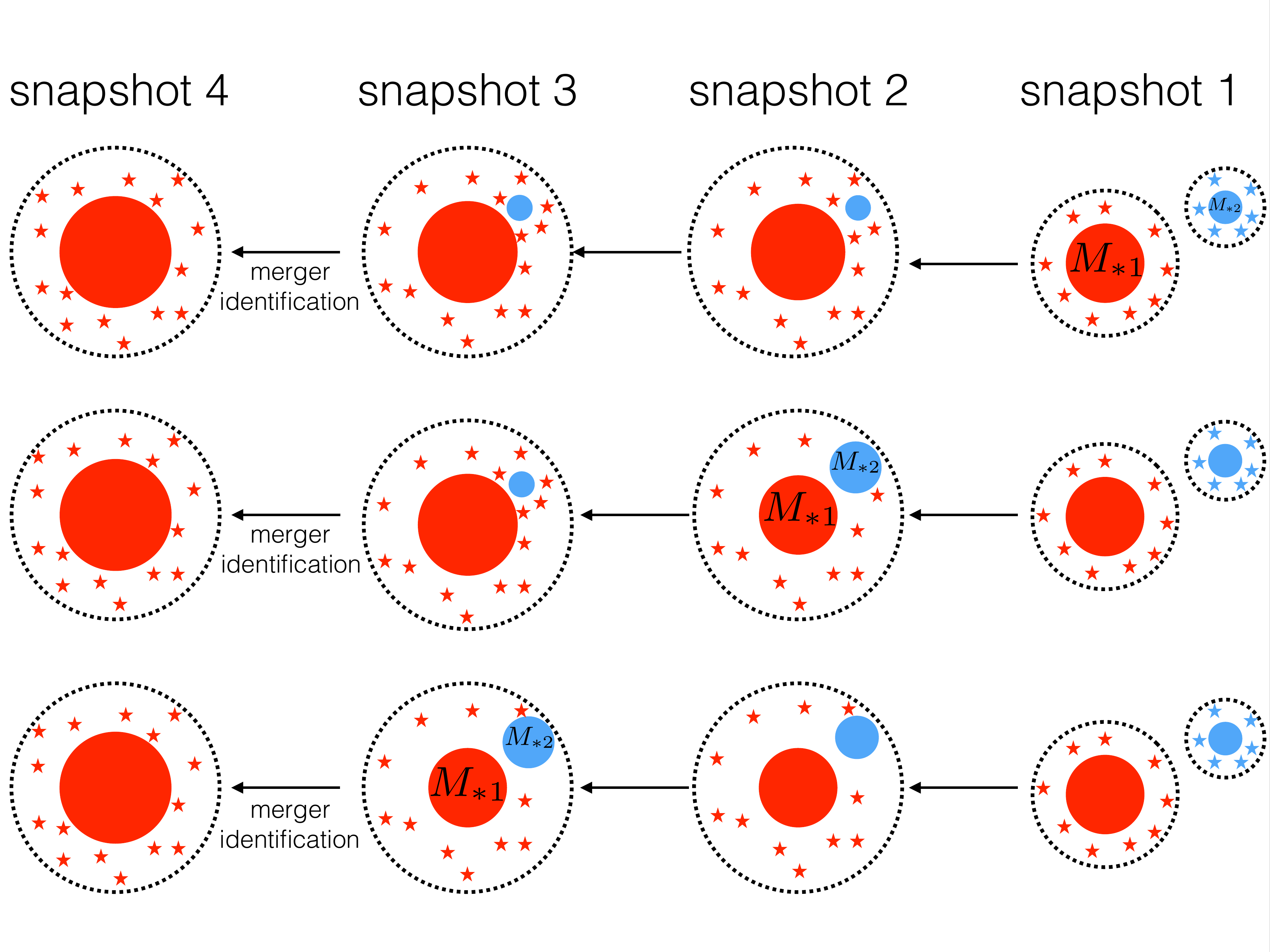}
  \caption{This sketch shows three different scenarios to illustrate of our definition of mergers and of the stellar merger mass ratio. The arrows show the direction of the time-line. The most massive galaxy is shown in red and the smaller progenitor is shown in blue. The filled circles show the galaxies and the dark matter halo is shown as dashed black line. The stars illustrate the intra-cluster light (ICL), which is always associated to the most massive galaxy within a dark matter halo. The size of the circles is associated to the stellar mass, which consists of the galaxy plus the ICL.
  Due to that definition of the stellar mass including the ICL and also to exclude effects like stellar stripping, the stellar masses in the last snapshot where \textsc{subfind} identifies two galaxies are no good proxy to estimate the stellar merger mass ratio.
  Thus, we trace the progenitor galaxies from the snapshot in which the merger was identified back until they were associated to different haloes. To estimate the stellar merger mass ratio we use the maximum mass of the second progenitor galaxy within all snapshots from the identification of the merger to the last snapshot in which they belonged to different haloes. In the three examples from top to bottom, the mass of the second progenitor galaxy is the largest in snapshot 1, 2, and 3, respectively.
	  }
\label{merger_definition}
\end{figure*}

In our simulations, the \textsc{subfind}-output is given in smaller time-steps than the snapshots of the simulation, which are mostly about 0.5~Gyr apart.
These snapshots are for example used to compute the gas parameters within the accretion radius.
Since the information about galaxy mergers is given by the \textsc{subfind}-output only, we can identify mergers also on smaller time-steps $t<0.5$Gyr.

In Fig. \ref{merger_definition} we illustrate our definition of galaxy mergers and how we estimate the stellar merger mass ratio, showing three different possible scenarios.
The most massive galaxy is shown as red filled circle and the less massive progenitor galaxy is shown as blue filled circle.
We know about the merger as soon as \textsc{subfind} identifies the two progenitor galaxies as separate subhaloes (snapshot 3 in our example).
These subhaloes can already be associated to the same dark matter halo, shown as black dashed circle.
Let us at first concentrate on the example shown in the bottom row to understand why choosing the stellar masses in the snapshot right before the merger of the subhaloes would lead to artificially small merger mass ratios:
\begin{itemize}
    \item{\textsc{subfind} associates the intra-cluster light (ICL, illustrated as stars) always to the most massive galaxy within a dark matter halo. Consequently, stars which originally belonged to the smaller progenitor galaxy (blue stars in the sketch) are associated to the larger galaxy as soon as the two progenitors are within the same dark matter halo (snapshots 2 and 3 in our example). Thus, the mass of the less massive galaxy would be underestimated and the mass of the more massive galaxy would be overestimated.}
    \item{In addition, it is possible that the two galaxies already interact. In that case effects like stellar stripping can also lead to an association of the stripped stars to the larger galaxy. Furthermore, some of the stars from the less massive galaxy might already have been accreted by the more massive one.}
\end{itemize}
To avoid these artificial problems, the merger mass ratio is generally computed before the two dark matter haloes merge (upper row in Fig. \ref{merger_definition}).
However, this is often long before the actual merger of the galaxies. Thus, between the merger of the dark matter haloes and the merger of the subhaloes, the galaxies might for example accrete or form a significant amount of stars.
Therefore, to further improve the method, we use the masses before the merger of the dark matter haloes only in cases, where the mass of the satellite galaxy is larger than afterwards. Therefore, we always trace the progenitor galaxies back to the last snapshot where they had separate dark matter haloes (snapshot 1 in our examples).
All in all we get the best estimate for the merger mass ratio when we choose the maximum stellar mass of the smaller progenitor galaxy within all snapshot from the identification of the merger to the last snapshot with separate dark matter haloes.
This might be, as generally assumed, before the merger of the dark matter haloes (upper row in Fig. \ref{merger_definition}), right before the identification of the galaxy merger with \textsc{subfind} (bottom row in Fig. \ref{merger_definition}), or in between (middle row in Fig. \ref{merger_definition}).

\bsp

\label{lastpage}

\end{document}